\documentclass[article ]{elsarticle}

\usepackage{lineno,hyperref}
\usepackage{xcolor}
\usepackage{amsmath,amsfonts,amsthm}
\usepackage{graphicx}
\usepackage{epstopdf}
\usepackage{mathtools}
\usepackage{subfig}
\usepackage{float}
\usepackage{bm}
\modulolinenumbers[5]
\newcommand{\erf}{\mathrm{erf}}
\newcommand{\e}{\mathrm{e}}
\newcommand{\D}{\mbox{d}}

\newcommand{\cH}{{\cal H}}

\newcommand{\bbbone}{{\mathbb{I}}}
\newcommand{\fbbone}   {{\mathbf{1\hspace{-2mm}1\hspace{-2mm}1}}}  

\newcommand{\rd} {\mathrm d}
\newcommand{\re} {\mathrm e}
\newcommand{\ri} {\mathrm i}

\newcommand{\nn}  {\nonumber} 
\newcommand{\be}  {\begin{equation}} 
\newcommand{\ee}  {\end{equation}}  
\newcommand{\bea} {\begin{eqnarray}}
\newcommand{\eea} {\end{eqnarray}} 
\newcommand{\ldd} {\Big\langle\!\!\Big\langle}
\newcommand{\rdd} {\Big\rangle\!\!\Big\rangle}

\journal{Physica A}
\begin{document}

\begin{frontmatter}

\title{Opinion dynamics with emergent collective memory: \\ the impact of a long and heterogeneous news history}

\author[1]{Gioia Boschi}
\ead{gioia.boschi@kcl.ac.uk}

\author[1,2]{Chiara Cammarota}
\ead{chiara.cammarota@uniroma1.it}

\author[1]{Reimer K\"uhn}
\ead{reimer.kuehn@kcl.ac.uk}

\address[1]{Mathematics Department, Strand, London WC2R 2LS, UK}
\address[2]{Dipartimento di Fisica, Sapienza Università di Roma, P.le A. Moro 5, 00185 Rome, Italy}

\begin{abstract}
In modern society people are being exposed to numerous information, with some of them being frequently repeated or more disruptive than others. In this paper we use a model of opinion dynamics to study how this news impact the society. In particular, our study aims to explain how the exposure of the society to certain events deeply change people's perception of the present and future. The evolution of opinions which we consider is influenced both by external information and the pressure of the society. The latter includes imitation, differentiation, homophily and its opposite, xenophobia. The combination of these ingredients gives rise to a collective memory effect, which is triggered by external information. In this paper we focus our attention on how this memory arises when the order of appearance of external news is random. We will show which characteristics a piece of news needs to have in order to be embedded in the society's memory. We will also provide an analytical way to measure how many information a society can remember when an extensive number of news items is presented. Finally we will show that, when a certain piece of news is present in the society's history, even a distorted version of it is sufficient to trigger the memory of the originally stored information.
\end{abstract}

\begin{keyword}
\texttt{Opinion dynamics}\sep \texttt{External Information} \sep \texttt{Collective memory} 
\MSC[2010] 00-01\sep  99-00
\end{keyword}

\end{frontmatter}

\section{Introduction}
Nowadays, thanks to the world-wide diffusion of digital technologies, people are bombarded by a myriad of different news. With the increasing popularity of social media, news items 
incessantly appear on all screens, triggering people's attention and influencing the way in which they think and act. Understanding how opinions and choices are affected by external events has thus become of particular interest for different research fields, such as sociology, politics and marketing. In the last years, physicists have proposed many mathematical models \cite{castellano2009statistical, flache2017models,  baronchelli2018emergence} to understand how humans behave and form their opinions, giving birth to a new research field called sociophysics. Some of the most famous works in this field concern the use of models borrowed from statistical mechanics, such as the Ising model to study consensus \cite{galam1982sociophysics, galam1991towards,  borghesi2007songs}. Others introduced a variety of different models such as the Voter model \cite{clifford1973model, holley1975ergodic} or the Majority rule model \cite{galam2002minority, martins2013building}. A part of these works is dedicated to the study of how the society reacts to external information \cite{sirbu2017opinion, martins2010mass} coming from a single source \cite{carletti2006make} or multiple ones \cite{quattrociocchi2014opinion}.

In our recent paper \cite{BOSCHI2020124909} we presented a model of a society where opinions are formed through the combined effect of mutual influence among agents and driving through multiple external events. The key ingredient of the model is the way in which agents interact: when two agents have often been in agreement, they tend to be in agreement also in the future and the same happens with disagreement. This behaviour is the result of the combination of different human tendencies largely studied and applied in different social models. Together with imitation, which is the main ingredient of all of the previously mentioned models, we consider differentiation \cite{mas2010individualization, mas2014cultural, sirbu2013opinion, sznajd2011phase, radillo2009axelrod, martins2010mass}, homophily (an attractive influence or the tendency to become more similar to people in agreement with us) \cite{axelrod1997dissemination, deffuant2000mixing} and its opposite, xenophobia (a repulsive influence or the tendency to become different from people in disagreement with us)  \cite{macy2003polarization, flache2011small, mark2003culture, huet2010openness, jager2005uniformity}. Our study shows that these combined attitudes, encoded in the interaction matrix of our model, prompt the emergence of a Hopfield-like \cite{hopfield1982neural, macy2003polarization} collective memory by which the society embeds the information coming from external news and recalls them in the future. Note that the memory effect which appears in our model is a memory of past {\em relations\/} and must not be confused with the memory of past actions or opinions of single agents that has been more often considered in the literature of social interactions \cite{dall2007effective, jkedrzejewski2018impact, stark2008decelerating}. Studies about collective memory appear instead in \cite{garcia2017memory}, while a model which consider memory of past relationships is presented in \cite{mariano2020hybrid}.

While the focus of our previous paper was on finite number of news cyclically presented, in this new work we study the behaviour of the same model but focusing our attention on items of information which are presented in random order and with non-uniform intensity. The exploration of this new scenario goes in the direction of making the model more realistic. In fact, in the real world, the news that a person may receive are plenty, with some of them appearing more often than others. We begin by exploring a simple scenario in which only a finite number of news items is presented, before moving on to more complicated ones in which information consists of extensively many items of news. In this second scenario we will use techniques common in statistical mechanics (i.e. replica calculations) to calculate the maximum number of pieces of information a society can retain in its collective memory. 

In that context we also investigate the question, how well a society is able to collectively recall a piece of information embedded in its collective memory at some point in the past, if exposed to a news item which is a more or less distorted version of that ancient memory, thus bearing only imperfect similarity with it.

We will present our results organizing the paper in the following way. In Sec. \ref{model} we present the model analysed in the paper. In Sec. \ref{sec:random} we present the behaviour of the society under the effect of a finite number of news items presented in a random order. In Sec. \ref{sec:fixed_cond} we show how the frequency and the strength of this news are determinant for their storage in the society's memory. In Sec. \ref{sec:random_cond} we perform a similar analysis testing a large spectrum of parameters. In Sec. \ref{sec:infinite_num_patt} we focus on a different setting in which many different news invest the society and we calculate how many among them will be effectively remembered, {\it i.e.} the storage capacity of our society, comparing analytical results with simulations. Finally in Sec. \ref{sec:noisy_signal} we show how a society which has received a certain signal, is able to remember it when the noisy version of the same signal is shown.

\section{The model}
\label{model}
 In this paper we will analyse a model,  firstly introduced in \cite{BOSCHI2020124909}, of a society of $N$ individuals interacting through feedback received from within the society itself and the effect of an external disruptive signal. In particular, with each agent we associate a continuous variable $u_i$  which represents her/his preference on a topic which evolves following
 \begin{eqnarray}\label{maineq}
\dot{u_i}=-u_i+I_i+\sum_{j(\neq i)}^N J_{ij}v_j+\eta_i\ .
\end{eqnarray}
In this model we assume that $v_j=g(u_j)$, representing the expressed opinions of the agent $j$, is a sigmoid (bounded) function of the preference field $u_j$. The dynamics of $u_i$ is driven by the agent's perception of external information $I_i$ and the pressure of the society, defined as the weighted sum of the expressed opinions of all the agents. The sign of the weights $J_{ij}$, which change in time  as described later, characterize the type of relation between the pair $(i,j)$ of agents: a positive sign entails an assimilation of opinions while a negative sign entails differentiation.
The time derivative $\dot{u_i}$ contains a mean reversion term $-u_i$ which entails that, when external influences are absent, the preference field of each agent fluctuates around zero. The last term $\eta_i$ is a white noise with Gaussian distribution with zero mean and finite variance $\langle \eta_i(t)\eta_j(t')\rangle = \sigma^2 \delta_{ij} \delta(t-t')$ which accounts for stochastic effects in opinion dynamics.

The novelty of our model is the way in which the couplings, and so the interpersonal relations of the agents', evolve in time. The tendency of two agents to agree or disagree is based on their history of agreement and disagreement. People which have a history of agreement (disagreement) will be more likely to agree (disagree) in future. This feature is encoded in couplings $J_{ij}$ which evolve in time following
 \begin{eqnarray}\label{eqJs}
J_{ij}(t)= \frac{J_0\cdot \gamma}{N} \int_{0}^{t}\mbox{d}s \ v_i(s)v_j(s) e^{-\gamma(t-s)} \ .
 \end{eqnarray} 
Here the product of the agents' expressed opinions is weighted with an exponentially decaying function, with a memory time scale $\tau_\gamma=1/\gamma$. 
It entails that more recent history has a larger weight than the distant past in determining the mutual influence of the agents.  In this way our society uses the past history to interpret the instantaneous inputs that receives from outside. 
The prefactor $J_0$ sets an overall scale for the strength of the interactions, while the $N^{-1}$ scaling of couplings with system size $N$ is to ensure that a large system limit of the dynamics exists.
A detailed study of the behaviour of this model under the influence of constant or periodic external information can be found in \cite{BOSCHI2020124909}. In the case of constant external information, we observed that at stationarity the $u_i$ are approximately Gaussian distributed $u_i \sim \mathcal{N}(\langle u_i \rangle, \sigma_u)$, with:
\begin{eqnarray}
\label{u}
\langle u_i\rangle &=& \sum_{j(\neq i)}J_{ij}\langle v_j\rangle + I_i\\
\label{g_4}
\sigma_u^2 &=&\sigma^2/2\ ,
\end{eqnarray}
and 
\begin{equation}
\label{g}
\langle v_i\rangle = \erf\left(\frac{\langle u_i\rangle}{\sqrt{1+2\sigma_u^2}}\right)
\end{equation}
where $\sigma^2$ is the noise variance. 

In this paper we will use the same Gaussian approximation to analyse the society under the effect of different kinds of random external signals. Firstly we will study the behaviour of the society under the influence of a finite number of different news items presented at random. Then we will study a society influenced by an infinite sequence of different news items and we will focus on how many of these can be actually remembered. 
 
\section{Random presentation of a finite number of patterns} 
\label{sec:random}
In our previous paper \cite{BOSCHI2020124909} on the model described by Eq.s (\ref{maineq}-\ref{eqJs}), we examined the society under the effect of sequential periodic external news. In that case the order of presentation determined which information was most clearly remembered. 
More realistically, however, we expect to see news appearing in a less regular fashion, with some signals more frequent or stronger than others. In this section we show how the society responds when news items may have a variable strength or are received in a random order, each returning with a given probability. 
This kind of external information will be represented in our model by a signal which is constructed from a collection of  many news items from which the  item presented at any time is chosen at random 
and is switched on for a time $\Delta_0$. Each piece of news is modelled by a random vector $\bm{\xi}^{\mu}$ with $\xi_i^\mu \in \{-1,1\}$ and an index $\mu \in \{1, \dots, p\}$. The value of $\xi_i^\mu$ determines the direction of the opinion that the agent $i$ takes in response to the information $\mu$. We can thus write the external signal as: 
\begin{eqnarray}\label{signal-4}
I_i(t) &=& I_{0}^{\mu_k} \xi_i^{\mu_k} \quad \mbox{for}\quad (k-1)\Delta_0 < t < k\Delta_0\ ,
\end{eqnarray}
where the $\mu_k \in \{1,\dots, p\}$ are random, with $\pi_\mu = \mbox{Prob}(\mu_k = \mu)$ defined as the probability that the information labelled $\mu$ is presented. When the $I_0^\mu$ are sufficiently large, the agents' opinions will quickly align with the information received, so that $v_i\simeq \xi_i^{\mu}$ almost immediately after the signal $\bm\xi^\mu$ is switched on. This corresponds to a situation in which the external information is disruptive and captures the attention of the whole society. In this case we are able to calculate the couplings $J_{ij}$, proceeding as in \cite{BOSCHI2020124909} for periodically presented patterns, splitting the integral over  intervals of length $\Delta_0$, during which $v_i(t) \sim \xi_i^{\mu}$. We adopt a  convention different from that used in \cite{BOSCHI2020124909}, counting ``backward" in time, and using $k=1$ to label the final presentation period, $k=2$ the previous one, and so forth. This gives
\begin{equation}
\label{J_random}
J_{ij}(t)=\frac{J_0}{N}\sum_{\mu=1}^{p} \xi_i^{\mu} \xi_j^{\mu} \, \sum_{k=1}^{N_p} w_k \,\delta_{\mu,\mu_k}\ ,
\end{equation}
with $N_p$ denoting the total number of news presentation periods up to time $t$, i.e. $t=N_p\Delta_0$, and
\begin{equation}
w_k = (1-{\rm e}^{-\gamma \Delta_0})\,{\rm e}^{-\gamma(k-1)\Delta_0}\ .
\end{equation}
The $k$-sum in Eq.\,\eqref{J_random} is a large weighted sum of Bernoulli  random variables $\delta_{\mu,\mu_k}$. It can be evaluated as a sum of averages by appeal to the Law of Large Numbers, giving
\begin{equation}\label{mean_J_random_long_time}
J_{ij}=\frac{J_0}{N}(1-{\rm e}^{-\gamma\Delta_0 N_p}) \sum_{\mu=1}^p \pi_\mu\xi_i^\mu\xi_j^\mu\ .
\end{equation}

We will be mostly dealing with the limit $\gamma\Delta_0 \ll 1 \ll \gamma\Delta_0 N_p$ for which $\sum_{k=1}^{N_p} w_k = 1-{\rm e}^{-\gamma\Delta_0 N_p} \simeq 1$, and for which it can be shown that corrections to Eq.\,\eqref{mean_J_random_long_time} are negligible, being smaller than the dominant contribution by a factor ${\cal O}(\sqrt{\gamma_0\Delta_0})$. If  $\pi_\mu=1/p$ for $\mu = 1, \dots, p$, couplings are (apart from a factor $1/p$) equal to Hebb-Hopfield couplings \cite{hebb1949organization, hopfield1982neural} used in the well known Hopfield model of  associative neural network. 
This similarity, already discussed in \cite{BOSCHI2020124909}, suggests that our network of agents can behave in a manner analogous to a network of neurons, in which firing patterns are stored in the couplings.
Instead of neural patterns we will have {\it opinion patterns} $\boldsymbol{\xi}^\mu$ that under suitable conditions can be stored and recalled by the society by means of a reshaping of individuals' interactions. A measure of the similarity between the stored opinion patterns and a given piece of news $\mu$ is the overlap between the pattern $\mu$ and the system state 
\begin{equation}
\label{m}
    m^\mu (t) = \frac{1}{N}\sum_i \xi^\mu_i \langle v_i(t)\rangle \ .
\end{equation}
The overlap $m^\mu$ will be close to one during the presentation of the signal contribution $\mu$ and close to zero when other signal contributions are presented. Most importantly, when the signal is totally removed, the value of $m^\mu$ tells us whether the society's opinions remain aligned with one of the previously presented news. 

To investigate this question, we evaluate the $m^\mu$ in a stationary regime with fixed external signal $I_i$, after couplings are frozen at their values given by Eq.\,\eqref{mean_J_random_long_time}. Exploiting the fact that the $u_i(t)$ in the stationary regime are Gaussian, with means and variances defined in terms of Eq.s \eqref{u}, \eqref{g_4} and \eqref{g}, we obtain
\begin{eqnarray}
\label{m2}
    m^\mu &=&\frac{1}{N}\sum_i\xi^\mu_i\, \erf\left(\frac{\langle u_i\rangle}{\sqrt{1+2\sigma_u^2}}\right) = \frac{1}{N} \sum_i\xi^\mu_i \erf\left(\frac { \sum_j J_{ij}\langle v_j\rangle}{\sqrt{1+\sigma^2}}\right)\nonumber \\
    &=& \frac{1}{N} \sum_i\xi^\mu_i \, \erf\left(\frac{J_0 \sum_\nu \xi_i^\nu \pi_\nu m^\nu + I_i}{\sqrt{1+\sigma^2}}\right)\ ,
\end{eqnarray}
i.e., a set of self-consistency equations for the overlaps $\{m^\mu\}$, $\mu= 1, \dots, p$. Here we have used the approximation $1-{\rm e}^{-\gamma\Delta_0 N_p} \simeq 1$ valid in the large $N_p$ limit as discussed above. For large $N$, the fixed point equations can be expressed in terms of averages over the statistics of the $\xi_i^\mu$ and the $I_i$ by appeal to the LLN, giving
\begin{eqnarray}
   m^\mu &=& \left\langle\!\!\left\langle \xi^\mu \erf\left(\frac{J_0 \sum_\nu \xi^\nu \pi_\nu m^\nu + I}{\sqrt{1+\sigma^2}}\right) \right\rangle\!\!\right\rangle_{\{\bm \xi^\mu\},I}\ ,
\end{eqnarray}
in which the double angled brackets denote an average over the distribution of the $\{\bm{\xi}^\mu\}$ and the external stationary signals $I$. In much of what follows we will be interested in so-called retrieval solutions in the absence of external signals, for which it is assumed that the system state is aligned with only a single opinion pattern, so that the vector $\mathbf{m}$ of overlaps has only a single non-vanishing component, $\mathbf{ m }=(0,...,m^\mu,...0)$, for some $\mu \in\{1, \dots , p\}$, for which the system of self-consistency equations \eqref{m} simplifies to
\begin{equation}\label{m^mu}
    m^\mu = \erf\left(\frac{J_0 \pi_\mu m^\mu}{\sqrt{1+\sigma^2}}\right)\ .
\end{equation}

The solution of Eq.\,\eqref{m^mu} can be compared to the value of $m^\mu$ estimated by simulating the model dynamics described in Eqs\,\eqref{maineq}, \eqref{eqJs}, \eqref{signal-4}, using presentation probabilities $\pi_\mu$ for the different patterns. Equations of motion are integrated using a simple Euler algorithm.
During the dynamics the couplings evolve and reach a plateau corresponding to their stationary value. When this happens we stop the dynamics and we freeze the couplings at their current value. 
We then switch on each signal pattern in turn for a brief period, sufficient to have the society aligned with the patterns, after which the signal is removed. We then wait until the value of $m^\mu$ converges to a ``stationary" level around which, due to the noise, it oscillates with a small variance. The final average value of $m^\mu$ is evaluated averaging over 700 values of $m^\mu(t)$ at stationarity. Results are then averaged over 10 simulations performed with different random pattern realizations. 
\begin{figure}
\centering
\includegraphics[scale=0.5]{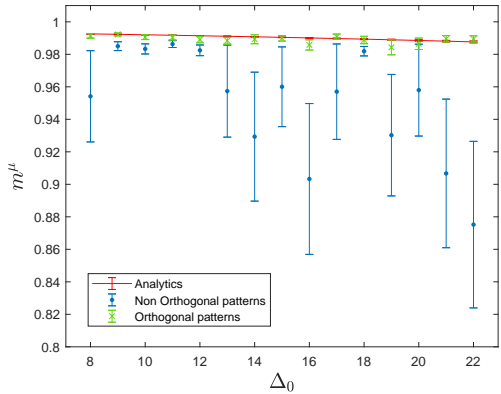}
\caption{The figure shows the comparison between the analytical (full line) and simulated (dots) value of $ m^\mu$ as a function of the presentation time length $\Delta_0$, for $J_0=6$, $I_0=10$ and $\pi_\mu=1/3 \ \ \forall \mu$.}\label{5}
\end{figure}
In the simulation we assume that $I_0^\mu=I_0$ and --- initially --- for $\pi_\mu=1/p$ for all $\mu$. In this way all the patterns have the same probability of appearance and the same strength. Other details of how simulations are performed can be found in \ref{app:methods}.
In Fig\,\ref{5} we compare the results for $m^\mu$ as a function of the presentation period $\Delta_0$ as obtained from simulations with the analytical predictions. For the simulations we use systems of size $N=100$, and $p=3$ for the number of patters which, as argued in \cite{BOSCHI2020124909}, is representative of a system at low loading. As $\Delta_0$ becomes large, results for random patterns begin to deviate from the theoretical predictions. This is mainly due to the effect that random patterns have non-zero mutual overlaps in systems of finite size $N$, which is in contrast to assumptions used in the derivation of the fixed-point equation for retrieval solutions in the thermodynamic limit $N\to \infty$. Indeed, using orthogonal patterns in the simulation, we observe a much better agreement between analytical predictions and results from simulations. To better understand this aspect we must carefully think of the role played by
non-orthogonality: given two positively correlated patterns, the memory of the first gets reinforced while the second is presented and vice versa. When two patterns are anti-correlated, the opposite happens and their memory is instead weakened. 
This phenomenon is amplified when $\Delta_0$ is large because the same patterns have more time to get mutually reinforced or weakened as the case may be. The creation of this pattern imbalance may lead to some patterns having a smaller overlap than others, or even being forgotten, which explains the deviation from predictions by numerical data for non-orthogonal patterns.

When the signal strength $I_0^\mu$ and the probability $\pi_\mu$ vary between patterns, the behaviour of the society is non trivially affected. The influence of varying these two parameters is investigated in the following by performing two different numerical experiments.
In the first experiment, described in Sec. \ref{sec:fixed_cond}, we show how the recovery of a piece of news depends on the probability of its appearance and on its strength in the news stream. 
In the second experiment, described in Sec. \ref{sec:random_cond}, we investigate the recovery of patterns when their parameters are chosen at random, such to observe how the probability of a piece of news to be embedded in the society depends on the properties of the {\em other\/} news also presented in the same dynamics.

\subsection{Storing random news with different strengths and frequencies of appearance}
\label{sec:fixed_cond}

In the previous experiment we analysed the behaviour of a society under the effect of news items presented randomly with equal probability. With the aim of modeling a more realistic scenario we consider now an external signal made of random news which have a different probability of being presented to the society as well as different intensities. To start this new analysis we will assign to the first pattern a probability $\pi_1$ and a signal strength $I_0^1$ and equal parameters to the other two with $\pi_2=\pi_3= (1-\pi_1)/2$ and $I_0^2=I^3_{0}$. In this experiment we will follow the same simulation protocols as before, except that we do not average the value of $m^\mu$ over several realizations. Rather than recording average overlaps, we measure the fraction of instances $f_\mu$ in which $m^\mu$ exceeds a critical value $m_c=0.4$. This value is chosen to be significantly larger than the value of random mutual overlaps between different patterns (see \ref{app:methods} for other simulations details.) 
\begin{figure}
		\centering
		 \subfloat[\label{subfig_f1}]{\includegraphics[scale=0.4]{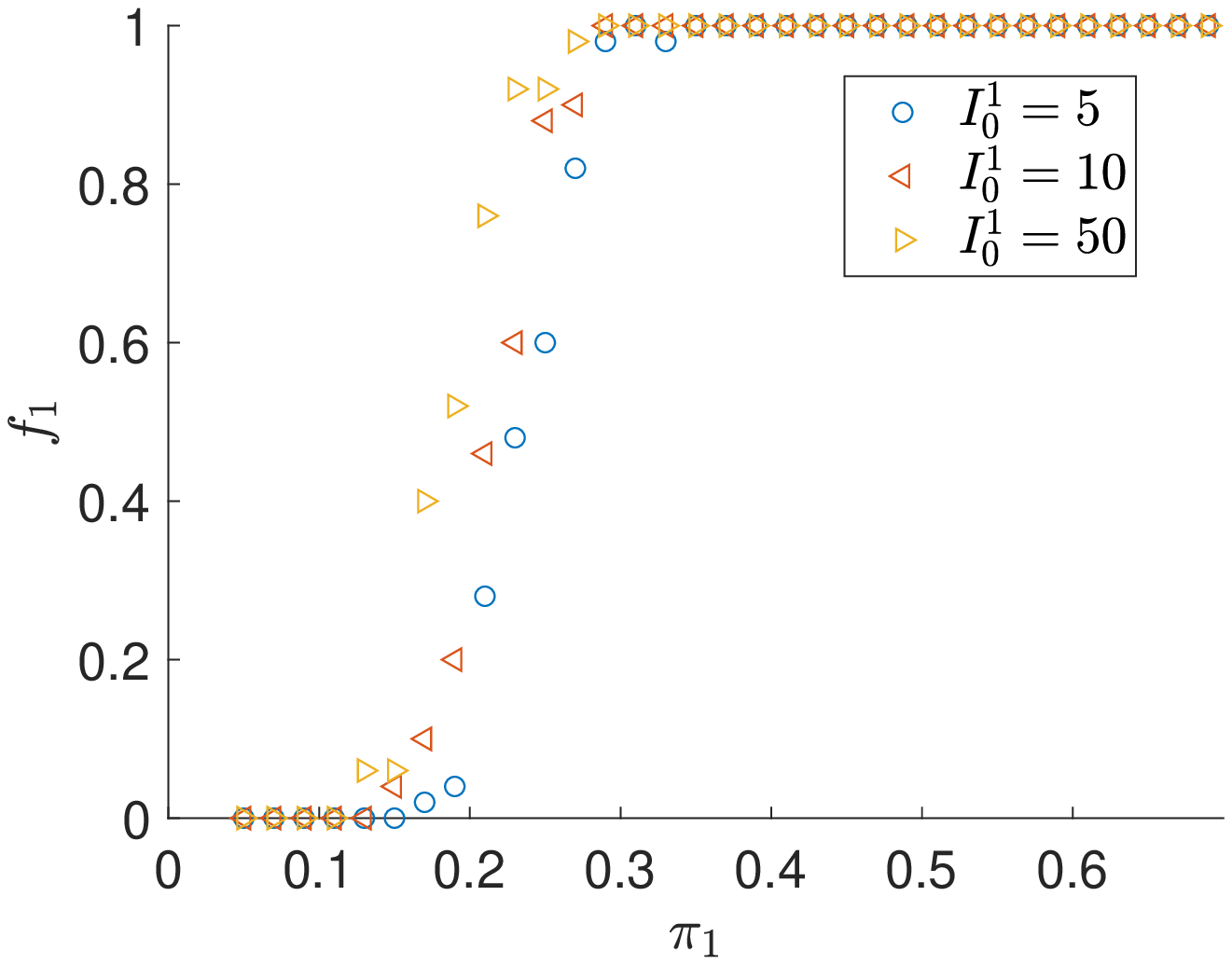}}
		  \subfloat[\label{subfig_f2}]{\includegraphics[scale=0.4]{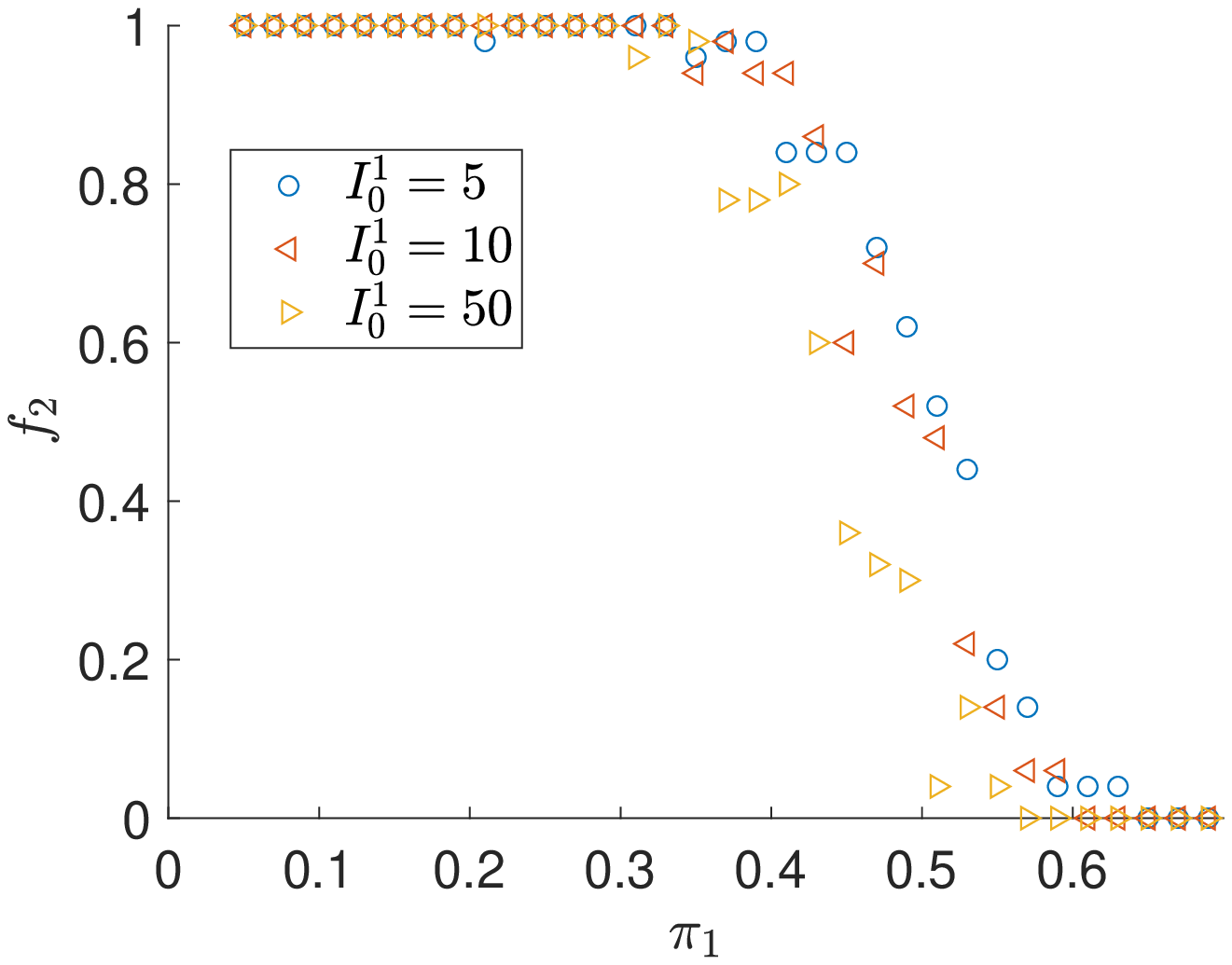}}
		\caption{(a) Fraction of instances in which pattern one is recovered as a function  of its probability of appearance.  (b) Same for pattern two. The quantities $f_1$ and $f_2$ have been calculated over 50 tests and then normalized on the interval $[0,1]$. The simulations have been performed using the following parameters: $I_0^2=I_0^3=10, J_0=8, \Delta_0=10$. }\label{23}
	\end{figure}
In Fig\,\ref{23} we show $f_1$ and $f_2$ as a function of $\pi_1$ for different values of $I_0^1$ keeping $I_0^2=10$ fixed. For all signal strengths, the fraction $f_1$ increases steeply as the pattern probability $\pi_1$ increases. We expect that for $N \to \infty$ the curve presents a  discontinuous transition between 0 and 1.
As anticipated, when the probability $\pi_1$ grows beyond 0.33 (the point where the three patterns are equally probable), the pattern $\mu=1$ is always recovered, while the other two are progressively less frequently remembered. However the most interesting effect is linked to the signal strength. In a real society we expect that the intensities $I_0^\mu$ of external information are not equal for all pieces of news but that some are significantly weaker or stronger than others. 
In Fig\,\ref{23} we show the effect of this variability in our model. The recovery of the first pattern becomes easier for $I_0^1>I_0^2 = I_0^3$  and more difficult for $I_0^1<I_0^{2,3}$. This result suggests that even a rarer piece of news, if particularly intense and disruptive, can leave a strong mark on the society.  

In Fig\,\ref{23} we can also notice that the increased strength of pattern one results in a more difficult recovery of pattern 2 even if its signal strength is not changed. This effect suggests that the possibility for a piece of news to impress the society depends not only on its own immediate impact but also on the impact of other news that is present in the news stream. This means that a piece of news strong enough to be remembered on its own can be forgotten when presented alongside stronger or more frequent information. This can explain, for example, how news about politics that would be remembered when broadcast alone may be forgotten when presented with other shocking news, such as those about terror-attacks or earthquakes.

\begin{figure}
    \centering
 \includegraphics[scale=0.4]{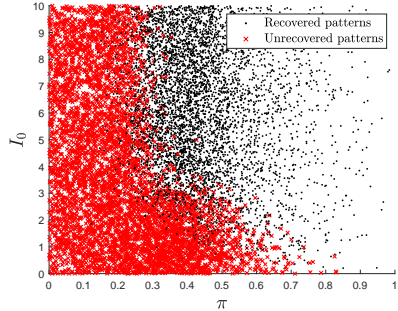} 
    \caption{ Black and red dots, corresponding to recovered and unrecovered patterns, are plotted as a function of their signal strength and probability of presentation. The parameters of the simulation are $N= 100$, $J_0=8$ and $\Delta_0=17$.}\label{fig_dots1}
\end{figure}

\subsection{Storing random news with random intensities and frequencies of appearance}
\label{sec:random_cond}

In the previous experiment we have shown that, when we fix the parameters of two pieces of information, the third one presented in the same experiment is more likely to be recovered if presented more often or if it is presented with bigger intensity. This gives us only a partial information on how our society reacts to random news. To complete this picture, we investigated the recovery of opinion patterns using a broad range of parameters. Using the protocol of Sec. \ref{sec:fixed_cond}, we now measure the recovery of patterns in a society where the signal strengths and probabilities of appearance are chosen at random, with  $0<\pi_\mu<1$, $\sum_{\mu=1}^p \pi_\mu =1$ and $0<I_0^\mu < 10$. As before we take $p=3$. The total number of samples of triplets taken is 3600. 
In Figure \ref{fig_dots1} we plot a map of recovered (black dots) and unrecovered (red crosses) patterns as a function of their probabilities of appearance and their strengths. For weak signals the patterns will not be remembered by the society even when they appear very often. For larger signal strengths, there is a separation between recovered and unrecovered patterns around $\pi=0.3$. 

We will call recovery zone the region of the graph where pattern recovery is predominant and will refer to the complementary region as non-recovery zone. 
However Fig\,\ref{fig_dots1} shows that some of the patterns around the boundary between the two zones don't behave as expected, e.g. some news around $I_0^1=5$ and $\pi=0.2$ are recovered, while some others around $I_0^1=5$ and $\pi=0.4$ are not.
More generally, the separation between the two zones is not sharp and we observe a large transition area. A smooth transition is to be expected due to finite size effects, as discussed earlier. 
In fact, we present patterns only for a finite number of times in any realization of the dynamics, giving rise to fluctuations in their effective frequency of appearance. However, other aspects may play a role, broadening even further the transition area visible in Fig\,\ref{fig_dots1}.
For instance as also discussed in the previous subsection the relative strength (and in this case the relative probability of appearances) of concomitant news in the recent history can result in a systematically more frequent recovery of news that would be otherwise unrecovered and vice versa. An attempt to disentangle these two aspects is presented in \ref{app:mutual_interaction}. 

\section{Extensively many patterns and society's storage capacity}
\label{sec:infinite_num_patt}
Real-life news are divulged with different intensity and characterized by different frequency of appearances. Moreover, it is also conceivable that the number of different news the society receives should not be considered to be very small in comparison to system size, as assumed when analyzing the situation of a finite number of external signals that we have considered so far.

We present here the analysis of a society hit by a stream of news items, each different from all others. We will take them as randomly generated at each presentation. In particular, we will investigate the number of news that the society can effectively remember, or the {\it storage capacity} of the society, whose interactions have been shaped by such a history of news. In order to store a large number of patterns, proportional to the number $N$ of agents of the system, the society needs to receive a number of news items of order $N$ {\em within\/} the memory time $\tau_{\gamma} = 1/\gamma$. For this reason we need to consider a scaling of the memory decay rate of the form $\gamma=\gamma_0/N$, as it is needed to ensure that an extensive number of patterns remains within ``memory range". In the initial formulation of the model (Eq.\,\eqref{eqJs}) we used an explicit scaling factor $1/N$ in the couplings in order to get a meaningful theory in the thermodynamic limit $N\to \infty$. In the present case, there is no need to introduce such an explicit scaling factor by hand, as the correct scaling follows automatically from the scaling of $\gamma$. Adopting such a scaling, the couplings $J_{ij}$ are thus seen to take the form
\begin{equation}
\label{Jij_infpatt1}
J_{ij}=\frac{J_0 \gamma_0}{N} \int_{0}^{t}\D s v_i(s)v_j(s) e^{-\gamma_0(t-s)/N}\ .
\end{equation} 
We should note that this scaling is correct in the limit of a large number of agents (thermodynamic limit), only when the society receives a series of different random news. We consider an external signal made of a sequence of news labelled by  $\mu = 1,2,3, \dots, p$, with $p = \alpha_{max} N$. Given that we are eventually considering a history of infinitely many patterns. i.e. $\alpha_{max} \to \infty$, it is convenient to count them in reverse order, with $\mu=1$ indicating the most recent pattern received and $\alpha_{max}N$ indicating the oldest one.

The time dependent external signal will thus be of the form
\begin{eqnarray}\label{signal-infinite}
I(s) &=& I_0 \xi_i^{\mu} \hspace{1cm} (\mu-1) \Delta_0 < t -s <  \mu \Delta_0\ .
\end{eqnarray}
As before, we assume signal intensities to be large. Each signal pattern $\mu$ is presented for a time $\Delta_0$, during which the society's opinions are well aligned with the signal with $v_i(s) \simeq \xi_i^\mu$. We can use this to evaluate the integral in Eq.\,\eqref{Jij_infpatt1}, splitting it in pieces of length $\Delta_0$ and calculating them separately as described in \cite{BOSCHI2020124909}. The resulting couplings for small $\gamma_0 \Delta_0$ are:
\begin{equation} 
\label{Jij_infpatt2}
J_{ij}=\frac{J_0 \tilde{\gamma}_0}{N} \sum_{\mu=1}^{\alpha_{max}N} \xi_i^\mu \xi_j^\mu e^{-\frac{(\mu-1)\tilde{\gamma}_0}{N}} + o(\tilde \gamma_0^2)\ .
\end{equation}
where 
we introduced $\tilde{\gamma}_0 \equiv \gamma_0\Delta_0$.
We will use $\alpha_c$ to denote the storage capacity of the society, i.e., $\alpha_c N$ is the maximum number of patterns that the society is able to recall. The concept of storage capacity was introduced in neural networks and first computed for the Hopfield model in \cite{amit1987statistical}. Our couplings in Eq.\,\eqref{Jij_infpatt2} are in fact reminiscent of a particular type of neural network model, the Hopfield model with forgetful memory \cite{nadal1986networks, parisi1986memory, van1988forgetful, marinari2019forgetting}. In the present paper, we investigate the storage capacity of a society in the noiseless limit of our model  \eqref{maineq}.
In this case, the dynamics of the $u_i$ is known to be governed by a Lyapunov function \cite{cohen1983absolute, hopfield1984neurons}, provided the interaction matrix is symmetric and the function $v_j$ describing the expressed opinions as functions of the preference fields $u_j$ are monotone increasing functions of their argument. Both conditions met in our case. Following  \cite{kuhn1991statistical, kuhn1993statistical} we can locate the minima of the Lyapunov function, and thus the attractors of the dynamics, by taking the zero-temperature limit of the free-energy of a system with the Lyapunov function as its energy function. The analysis can be found in \ref{app:replica}. We will describe the collective properties of the society using three order parameters:
\begin{enumerate}
\item the overlap between the 
society state and a given opinion pattern presented by the signal (taken to be pattern 
$\mu$) 
\be
m=\frac{1}{N} \sum_i \xi_i^\mu\, \overline{\langle v_i\rangle} \ ,
\ee
where angled brackets denote a thermal average and over-bars an average over the disorder embodied by the other opinion patterns $\nu (\neq \mu)$ embedded in the society. For a finite number of patterns, this parameters corresponds to the one defined in Eq.\,\eqref{m}.

\item the mean of averaged squared opinions, which are the off-diagonal elements of the replica symmetric matrix of Edwards-Anderson order 
parameters (see \ref{app:replica})
\be
q=\frac{1}{N} \sum_i \overline{\langle v_i\rangle^2} \ ,
\ee
\item a susceptibility-type parameter
\be
C=\beta(q_d-q)= \frac{\beta}{N} \sum_i \Big(\overline{\langle v_i^2\rangle}-
\overline{\langle v_i\rangle^2}\Big) \ ,
\ee
where $q_q$ are the diagonal elements of the replica symmetric matrix of Edwards-Anderson order 
parameters.
\end{enumerate}
 Making use of replica theory (details of the calculations in \ref{app:replica}) we obtain the following three fixed point equations, the solutions of which self-consistently determine these three order parameters and their dependence on the parameters characterising the system:
\bea
m &=&\ldd \xi^\mu\ \hat v \rdd\ ,\nn\\
C &=& \frac{1}{J_0\sqrt{r}} \ldd z \hat v \rdd \ ,
\label{fpes}
\\
q &=& \ldd \hat v^2 \rdd\ ,\nn
\eea
where
\be \label{v_hat}
\hat v(\xi^\mu,I,z) =g\bigg(m \xi^\mu J_0\tilde{\gamma}_0e^{-\tilde{\gamma}_0 \alpha} + J_0 \sqrt{r}z + I -  J_0 \Big(1 + \frac{\ln(1-J_0\tilde{\gamma}_0 C)}{J_0\tilde{\gamma}_0 C}\Big) \hat v\bigg )\ ,
\ee
with 
\be
r=\frac{q}{J_0C}\left[\frac{1}{1-J_0\tilde{\gamma}_0C}+\frac{\ln(1-J_0\tilde{\gamma}_0C)}{J_0\tilde{\gamma}_0C} \right]\ ,
\label{hatnu}
\ee
and where $z$ is a normally distributed random variable of zero-mean and unit variance. The double angled brackets denote an average over the site-random variables, i.e. $\xi^\mu$ and $I$, and over the Gaussian random variable $z$. We note that the equations above have been obtained under the assumption that during every presentation time the society is perfectly aligned with the signal patterns presented i.e. in the large $I_0$ limit. The $I$ appearing in Eq.\,\eqref{v_hat} is not the signal described by Eq.\,\eqref{signal-infinite} that appeared during the history of pattern presentations to the society, but a node-dependent perceived signal to which agents of the society may be subjected at the end of an extensive number of such pattern presentations. 
In the following subsections we will study two different retrieval scenarios, one is the spontaneous retrieval of patterns $\bm{\xi}^\mu$ obtained setting $I$ to 0, the other is the retrieval of old patterns when one of them is presented again, though in the form of a weak randomly distorted version of the original.

In both cases the problem of solving $z$-dependent fixed-point equations within the set \eqref{fpes} of fixed-point equations is avoided by transforming the Gaussian $z$- distribution into a $\hat v$-distribution as done in \cite{kuhn1993statistical} and then taking the averages respect to the $\hat{v}$-distribution. 

 \begin{figure}
\subfloat[\label{fig_inf_patt_2}]{\includegraphics[scale=0.4]{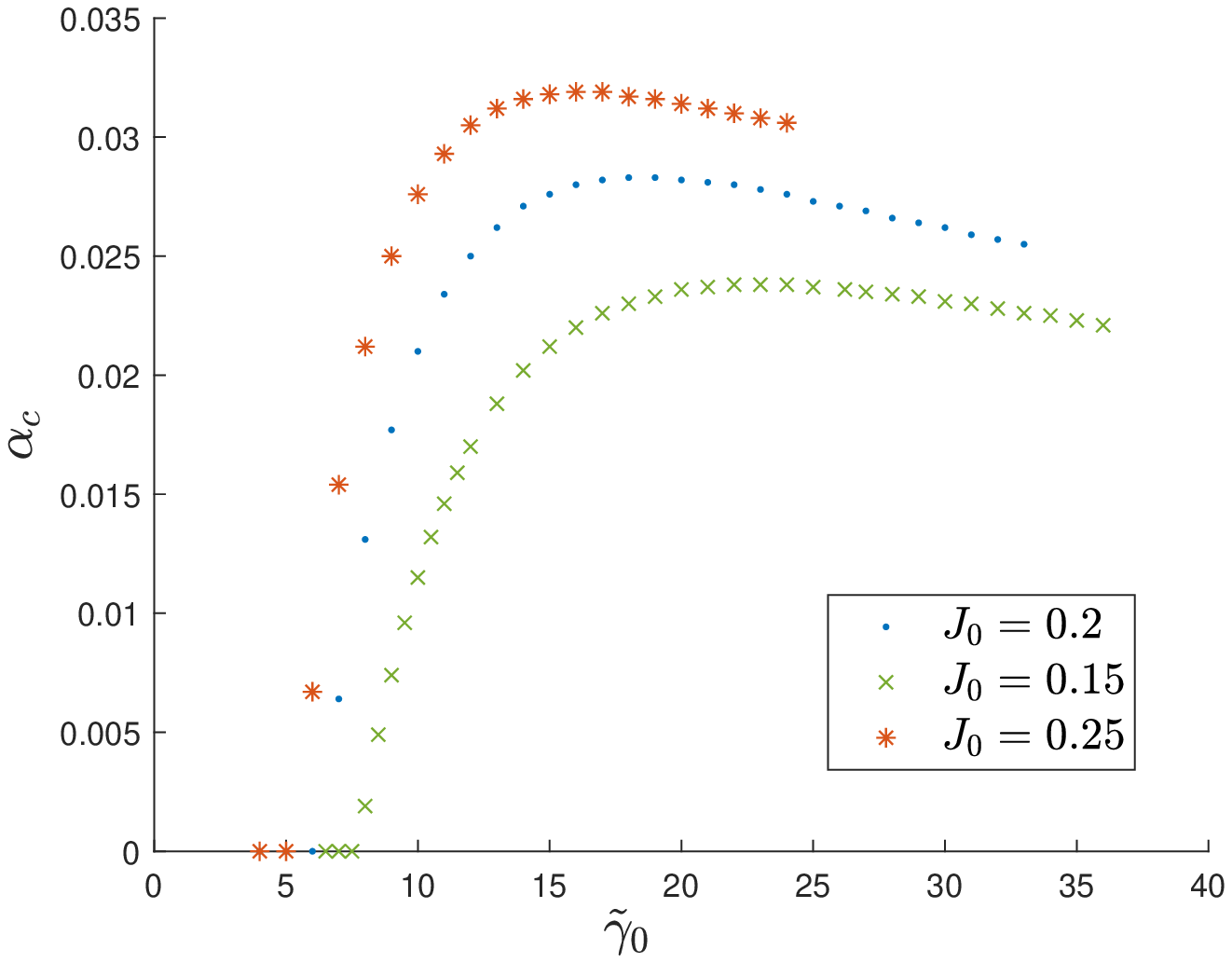}}
\subfloat[\label{fig_inf_patt_1}]{\includegraphics[scale=0.4]{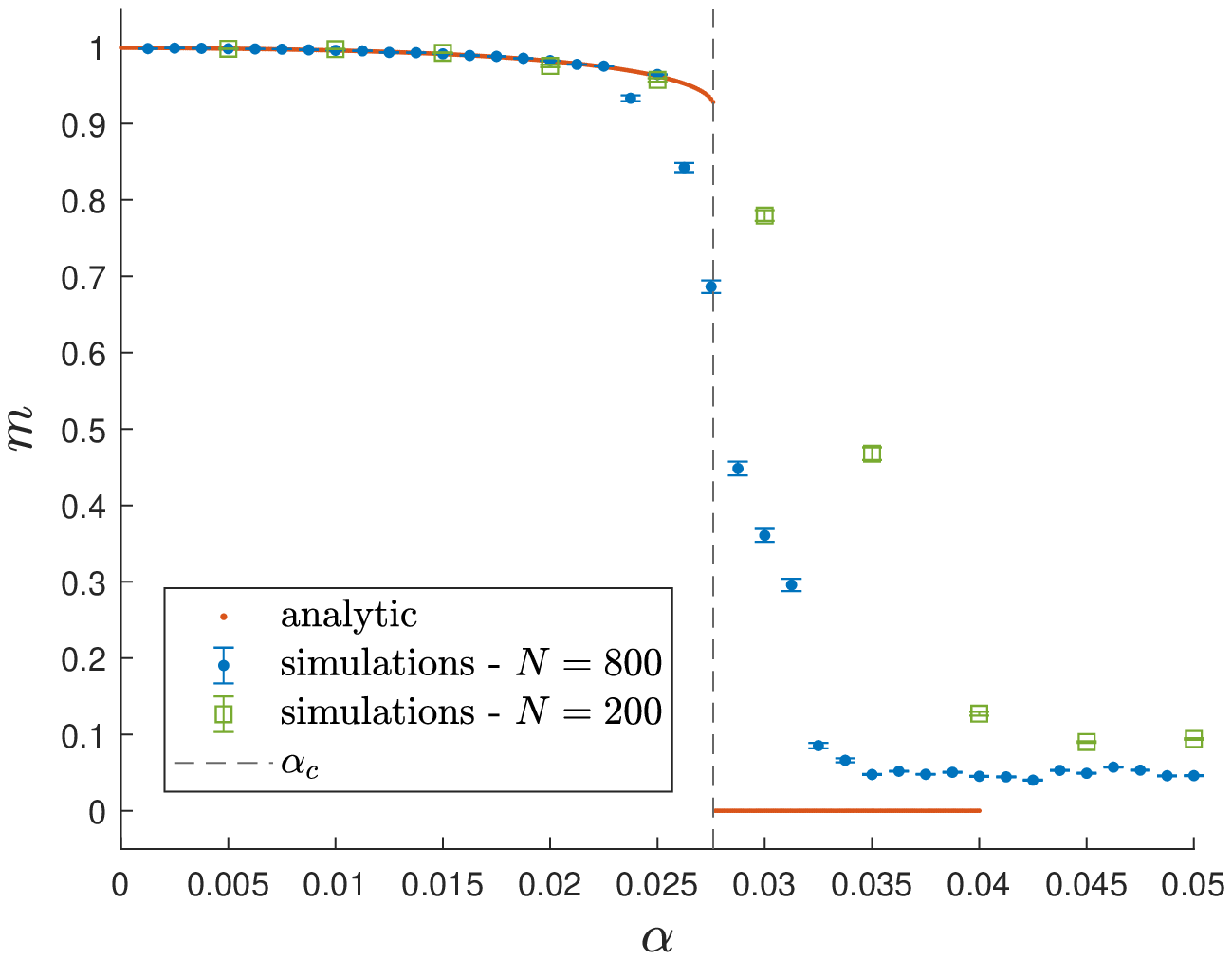}}
 \caption{Spontaneous retrieval of patterns in a society with an infinite history of news. (a) Critical capacity of the society as a function of $\tilde{\gamma}_0$ for different values of $J_0$. (b) Comparison between analytical solutions of $m$ for different values of $\alpha$ at $\tilde{\gamma}_0 = 15$ and $J_0 = 0.2$ and simulations of societies of different sizes with Gaussian noise with mean 0 and variance $\sigma^2 = 0.01$. The dashed line corresponds to the analytical critical capacity.}
     \label{fig:inf_patterns}
\end{figure}
\subsection{Spontaneous retrieval}
\label{sec:spont_retrieval}
We solved Eq.s \eqref{fpes}  numerically for $I=0$ and different values of $\alpha$ and  $\tilde{\gamma}_0$. 
We expect this system to have two solutions when $\alpha \leq \alpha_c$, one with $m=0$ and one with $m > 0 $, the latter disappearing for $\alpha > \alpha_c$. We thus computed the storage capacity $\alpha_c$ as the threshold of $\alpha$ after which the system has a unique solution with $m = 0$.
In Fig\,\ref{fig_inf_patt_2} the values of $\alpha_c$ found for different values of $J_0$ are plotted as a function of $\tilde{\gamma}_0$.  
 Whatever the value of $J_0$, the storage capacity is zero when $\tilde{\gamma}_0$ is too small; it becomes non-zero at a critical value of $\tilde{\gamma}_0$ and grows until it reaches a peak and then decreases again. A small value of $\tilde{\gamma}_0$ corresponds to a long memory time or very short pattern presentation periods. This results in a society being exposed to too many patterns within its memory time, which interfere with each other and make the retrieval of these memories impossible. Conversely, when $\tilde{\gamma}_0$ is too large, the memory of the patterns fades too quickly for a large number of patterns to be remembered. These phenomena are similar to those observed in \cite{nadal1986networks, parisi1986memory, van1988forgetful, marinari2019forgetting} in the context of forgetful Hopfield networks.
 
In Fig\,\ref{fig_inf_patt_1} we show an example of how the solution jumps from $m \sim 1$ to $m = 0$ when $\alpha > \alpha_c$. The numerical solution is compared to the values of $m$ found using the simulated dynamics at low temperature for the model described in Eq.\,\eqref{maineq}. In the simulations we present a sequence of $p = N$ signal patterns to a society of $N=800$ and $N=200$ individuals. We record the values of $m$ in absence of signal as described at the end of Sec. \ref{sec:random} and we averaged them over 50 simulations with different patterns realizations. Other simulation details can be found in \ref{app:methods}. Fig\,\ref{fig_inf_patt_1} shows the values obtained as a function of $\alpha = \mu/N$, which are in good agreement with the theory. Given that we have a finite $N$, the transition from one solution to the other at $\alpha = \alpha_c$ is not abrupt, but is rounded and becomes steeper with increasing $N$. For the same reason there is a non-zero overlap of order $\mathcal{O}(1/\sqrt{N})$ between the patterns, which results in a residual $m$ for $\alpha >\alpha_c$.   

\subsection{Noisy signal}
\label{sec:noisy_signal}
In this last subsection we study the ability of a society to retrieve patterns of news after having been exposed to an infinite series of such news items.
In particular, we focus on what happens when one of these pieces of news is presented again to the society, albeit at low intensity and distorted by some amount of noise. How will the society respond? Will it develop a reaction in line with the new pattern {\em as presented}, or will the interpersonal relations formed in response to the society's history of news exposures allow it to recognize and retrieve the information in its ``pure'' form as previously stored?
The answer depends not only on the strength of the signal presented and on how much it is distorted by the noise, but also on how long ago in the past the society had been exposed to the original un-distorted version of it. In other words, it depends on how well the memory of the original opinion pattern is still embedded in the system. In order to model this situation, we consider a noisy signal of the form:
\be\label{distorted_signal}
I_i = \widetilde{I_0}\xi_i^\mu + \sigma_I z_i
\ee
where $\xi^\mu$ is one of the news presented in the society's history, $\widetilde{I_0}$ the amplitude of the new signal to be presented and $\sigma_I z_i$ is a Gaussian noise with mean 0 and variance $\sigma_I$. 
\begin{figure}
    \centering
    \subfloat[\label{fig_corrupted_1}]{\includegraphics[scale=0.4]{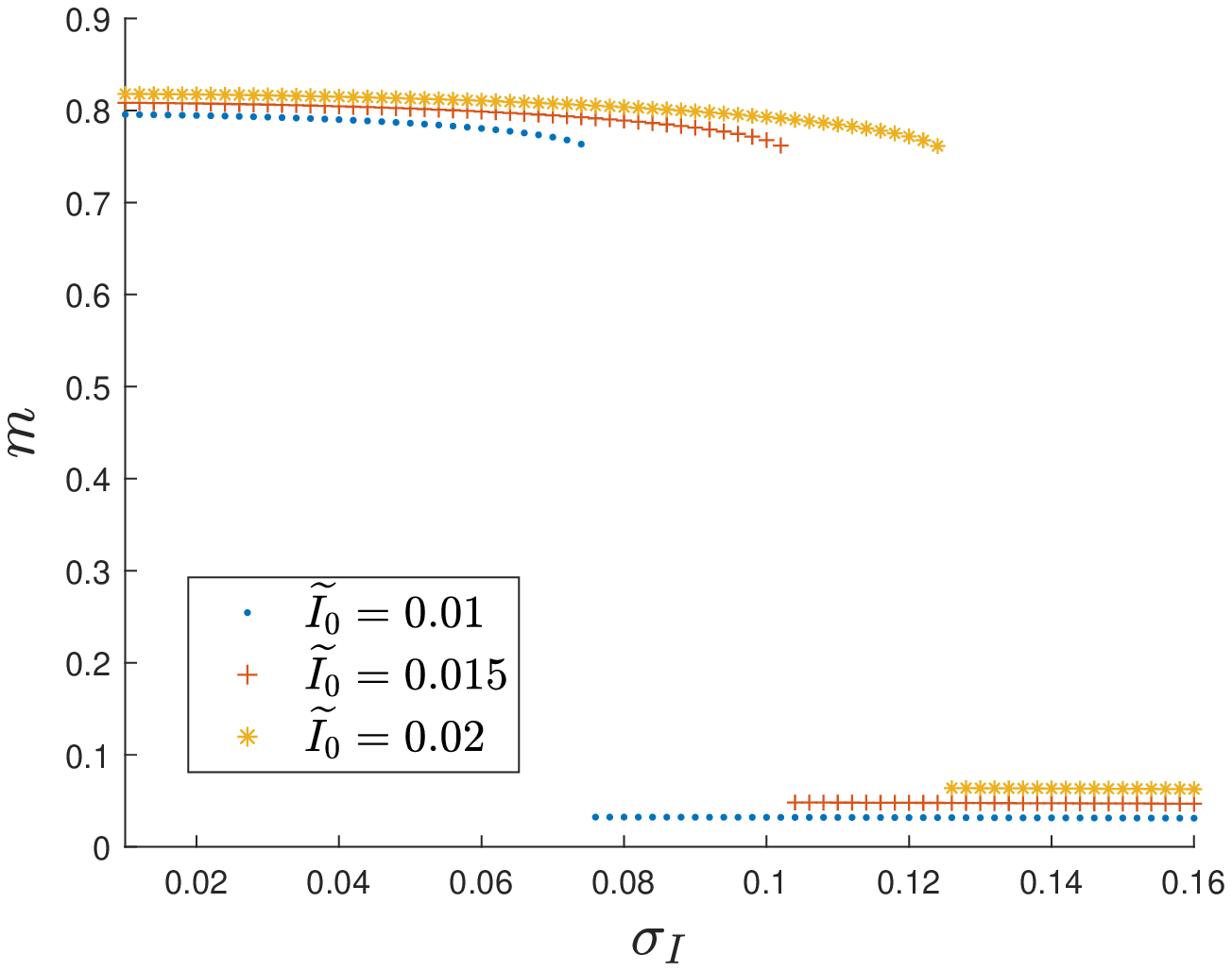}}
\subfloat[\label{fig_corrupted_2}]{\includegraphics[scale=0.4]{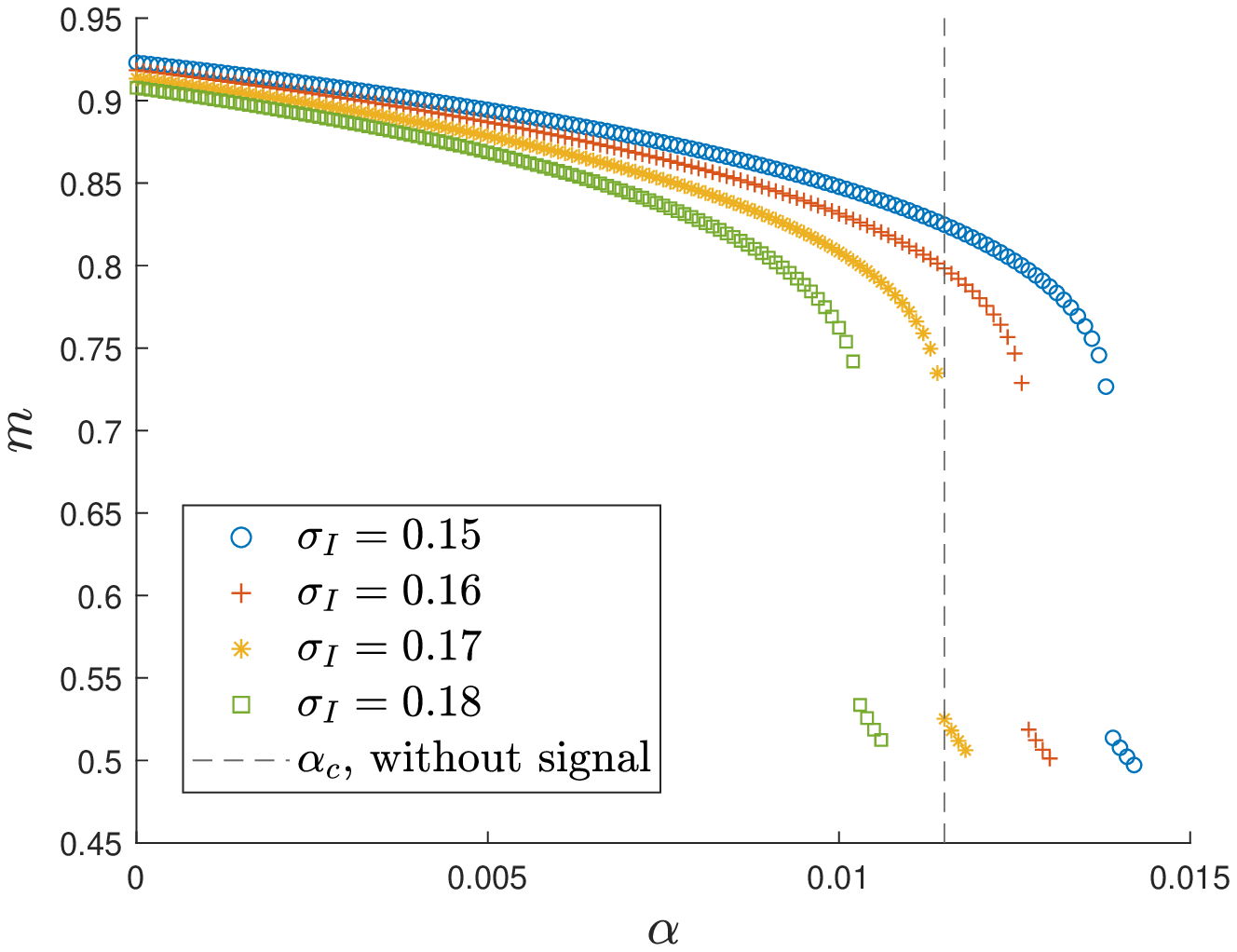}}\\
\subfloat[\label{fig_corrupted_3}]{\includegraphics[scale=0.4]{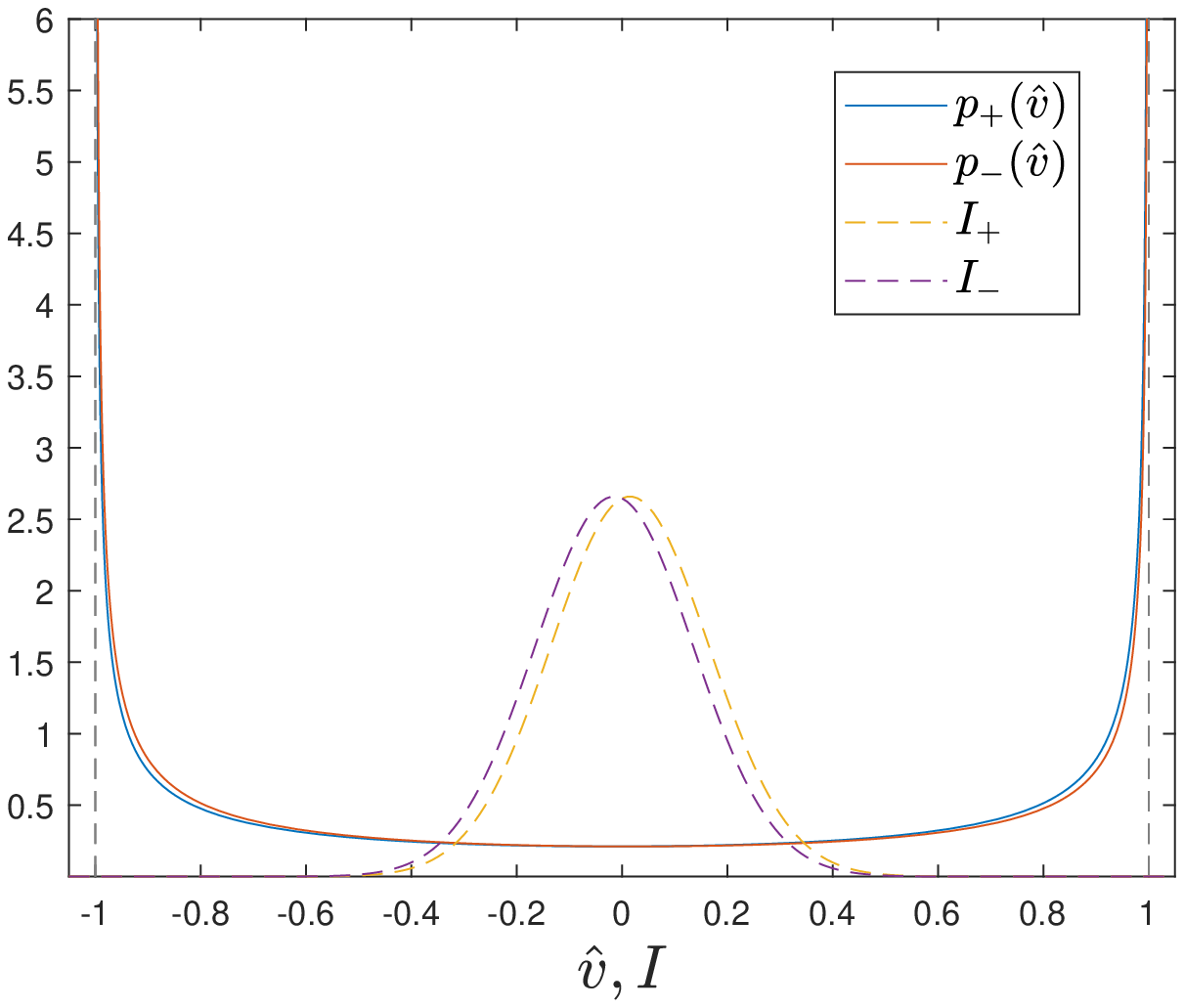}}
\subfloat[\label{fig_corrupted_4}]{\includegraphics[scale=0.4]{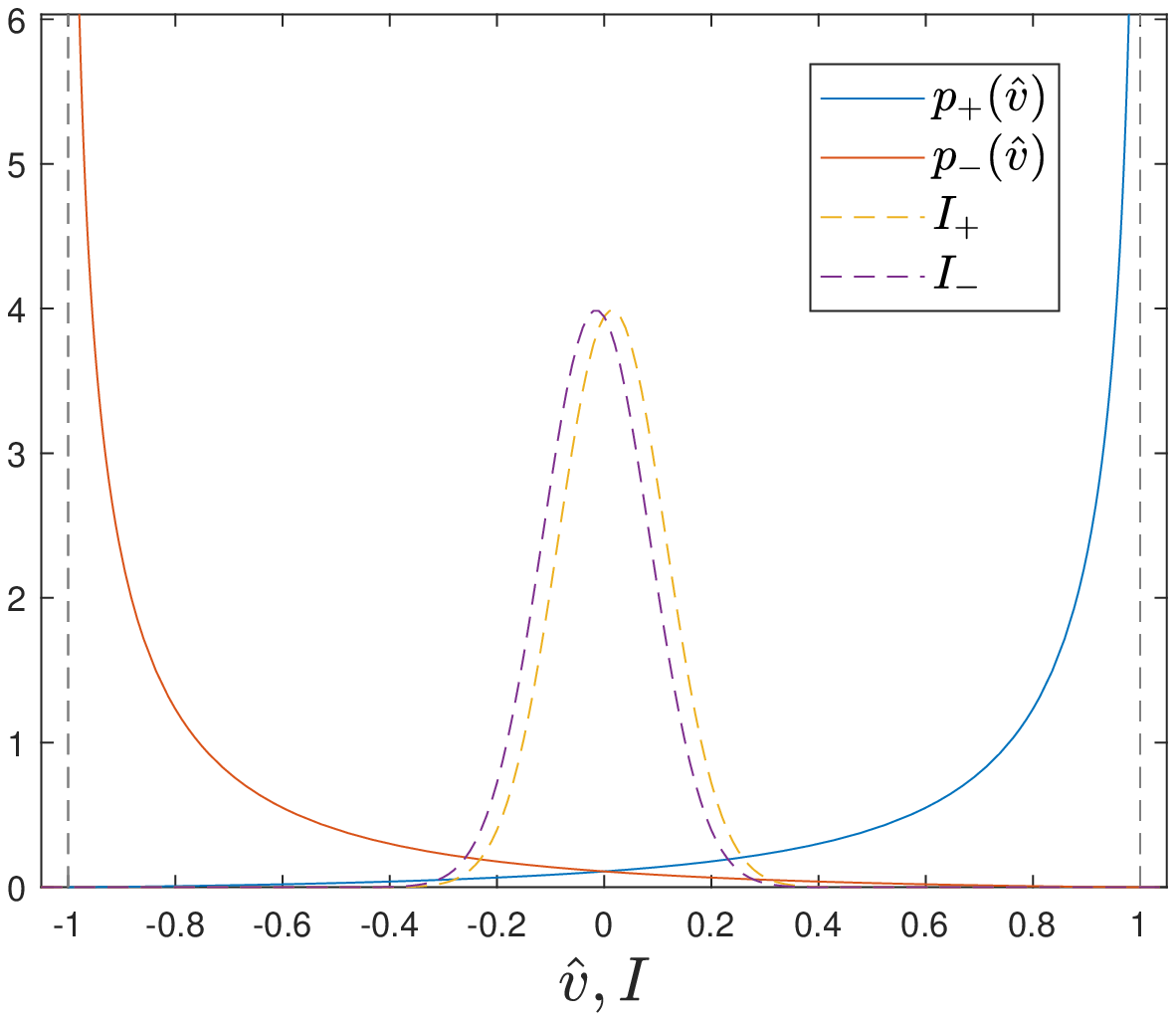}}
    \caption{Recovery of patterns through a noisy signal in a society which has received an infinite history of news. (a) Solution of $m$ as a function of the level of noise of the applied signal, for different values of $\widetilde{I_0}$, given $J_0 = 0.15$, $\tilde{\gamma}_0 = 10$ and $\alpha = 0.012$. (b) Solution of $m$ as a function of $\alpha$ for different level of noise and fixed $\widetilde{I_0} = 0.2$, $J_0 = 0.15$ and $\tilde{\gamma}_0 = 10$. The dashed line indicates the value of $\alpha_c$ in absence of the noisy signal. (c) Plot of the distribution of expressed opinions $p_{\pm}(\hat{v})$ (corresponding to $p(\hat{v})$ at $\xi^\mu = \pm 1$), for $\alpha = 0.012$, $J_0 = 0.15$, $\widetilde{I_0} = 0.015$ and $\sigma_I = 0.15$. The dashed lines correspond to the plots of the the noisy signals with average $\pm \widetilde{I_0}$. (d) Same plot with $\sigma_I = 0.1$.}
    \label{fig:corrupted_signal}
\end{figure} 

First we comment on the average behaviour of the society. Figure \ref{fig:corrupted_signal} shows the overlap $m$ with the original pattern for different values of $\sigma_I$ and $\widetilde{I_0}$ at $\tilde{\gamma}_0 = 10$ and $\alpha = 0.012$. In panel (a) it is visible how the retrieval of the original opinion pattern becomes more difficult as $\sigma_I$ grows, until the pattern is no longer recovered. Naturally, as $\widetilde{I_0}$ gets bigger, the point at which this happens moves towards larger values of $\sigma_I$. Conversely, when the noise is too disruptive the pattern is mostly not retrieved but the presence of the signal still guarantees a non zero value of $m$.

We show in panel (b) how the distortion of the signal also changes the maximal `age' $\alpha_c$ of a news item that the society is able to retrieve when exposed to a distorted version of it. While we use the same symbol as for the storage capacity, i.e. the maximum number of opinion patterns a society can {\em spontaneously\/} retrieve, these quantities are obviously not strictly the same. 
In particular in absence of distortion, the presence of a signal aligned with one news already seen produces a larger $\alpha_c$ than the storage capacity of spontaneous retrieval. 
In other words, the presence of $\widetilde{I_0}$ allows the retrieval of opinion patterns that would have not been recovered spontaneously. When some amount of distortion is also present, how far in the past these opinion patterns can have been presented and still be recalled, clearly depends on the level $\sigma_I$ of distortion of the pure memories. Figure \ref{fig_corrupted_2} shows how the value of $\alpha_c$ at which the solution for $m$ jumps between a high and a low value, depends on $\sigma_I$ when the distortion is added to a signal of strength $\widetilde{I_0}=0.2$. When the noise increases, the degree of similarity with an item originally stored decreases implying that older opinion patterns, which are less strongly embedded, cannot be recalled any longer; this is confirmed by the fact that the corresponding $\alpha_c$ decreases with increasing level $\sigma_I$ of pattern distortion. \\

These results on the average behaviour can be complemented by looking at the distribution of expressed opinions, which reveal the emergence of an interesting behaviour.
Fig\,\ref{fig_corrupted_3} and \ref{fig_corrupted_4} show how the distributions of expressed opinions
in response to a distorted incoming piece of news, change in a very non trivial way with its noise level $\sigma_I$. The functions $I_+$ and $I_-$ show the distribution of the external fields corresponding to positive and negative values of $\xi_i^\mu$, with
\be
I_\pm = \pm \widetilde{I_0} + \sigma_I z_i\ .
\ee
The function $p_+(\hat{v})$ is the distribution of expressed opinions of all agents $i$ which originally received the signal $I = I_0\xi_i^\mu$ with $\xi_i^\mu = +1$, while  $p_-(\hat{v})$ is the same for $\xi_i^\mu = -1$.
When expressed opinions have the same sign as the original signal we can conclude that the society has successfully retrieved the corresponding news.

Panels (c) and (d) show two scenarios in which the signal is in both cases very distorted and the distributions of $I_+$ and $I_-$ barely differ, being just slightly shifted versions of each other, yet the response of the society is very different. 
Panel (c) shows that in presence of a very large distortion
$p_+(\hat{v})$ and $p_-(\hat{v})$ are approximately symmetrical and almost identical. This means that the society is not able to retrieve the original opinion pattern and individuals' opinions are {\em uncorrelated\/} with the original un-distorted opinion pattern the society had been exposed to in the past.

A small reduction of the noise level or the amount of distortion of an original news item
leads to a completely different behaviour in terms of the distributions of expressed opinions. The results corresponding to this scenario are shown in panel (d), with distributions of expressed opinions being strongly asymmetric and biased in the direction of the original {\em un-distorted\/} signal with, $p_+(\hat{v})$ ($p_-(\hat{v})$) peaked at $1$ ($-1$) only. In this case individual's opinions are almost perfectly aligned with the originally presented information, despite the fact that the society is exposed to an external signal which has very little resemblance with it.

The important message delivered by these results is that even a weak and distorted signal can trigger the society to retrieve old stored memories. The society described by our model is able to remember information from the past, even when exposed to very noisy versions of it. The collective memory that emerges in this context connects distorted pieces of information to their clear un-distorted versions embedded in the collective memory. The maximum amount of noise tolerated for this to happen depends on the signal strength $\widetilde{I_0}$ and the value of $\alpha$, {\it i.e.} how far in the past an original news item has impacted the society. In a real society shaped by a history of intense events, like terror attacks, earthquakes, or other crises, a new signal which resembles one of the old memories, even if only barely, can trigger in the population the same reactions it had shown in the past. These results open the way to design social experiments to understand how collective memory is formed and activated or re-activated by new information, and to test whether this mechanism can explain strong and pronounced collective responses to news items that don't appear to be particularly disruptive. 

 \section{Conclusions}
In this paper we studied how exposure to strong external information can shape the behaviour of a society within a model of opinion dynamics introduced in our previous paper \cite{BOSCHI2020124909} that produces interactions between individual agents based on their recent history of mutual agreement or disagreement. 
In this way people tend to agree with others if they have a history of predominant of mutual agreement, whereas they will be more likely to disagree with them in case of a history of predominant recent disagreement. This mechanism gives rise to a collective memory effect by which a society can remember past configurations of opinions. As mentioned initially, the opinion configurations we are interested in are those produced in reaction to external news. In the present paper we studied the effect of random news and how properties, such as their frequency of appearance, their strength or their relative intensity with respect to other news, determine whether the corresponding opinion patterns can be remembered or not. Unsurprisingly, we have seen that opinions in reaction to strong and frequent news are more easily remembered. However, we observe that even rare signals can have a deep impact on the society if they are powerful enough. Moreover, we showed that the memory formation of an opinion pattern depends not only on its own strength and frequency of appearance, but also on the characteristics of the other news presented within the memory span of a society. In fact, a piece of news which has a sufficient strength  to be remembered if presented along with news items which have a similar amplitude, can be forgotten if were instead presented along with stronger news. A different aspect of the model behaviour is the proportion of consecutive news that a society can remember when a very long sequence of information is presented. Using techniques borrowed from statistical mechanics, we found this proportion analytically and confirmed it using simulations. The results which we found are compatible with the behaviour of models appearing in the literature of forgetful Hopfield networks \cite{nadal1986networks, parisi1986memory, van1988forgetful, marinari2019forgetting}. The last result concerns how noisy information is perceived by the society. We showed that if a piece of news was presented in the recent history of a society, and it is confronted with distorted version of it, the society is able to reconnect the distorted version with the clean memory it corresponds to, recalling the originally stored memory of it. This implies that the collective memory of a society is able to produce interesting effects that modulate the collective perception of current news or affairs, and it could be used to understand attitudes of people towards external events that may be only weakly correlated with past ones. This means that a shocking event that hits a society can change its perception of future similar events.

In the future we would like to extend our work by adding other interesting features to our model. We know for example that in the real world not all people receive the same news or perceive them with the same strength, so we aim to take this into account. Another possible improvement consists in personalising the intensity of the pressure of the society for different agents. In this way some people will be more influenced by their peers and others less. Another possibility is to consider multiple opinions related to a set of different issues that may be relevant at the same time, and to have interactions depending on the history of such sets of opinions on a range of different topics. Apart from the addition of further ingredients to the model, the main future direction of our work will be its calibration against real data. 

\section*{Acknowledgements}
The authors acknowledge funding by the Engineering and Physical Sciences Research Council  (EPSRC)  through  the  Centre  for  Doctoral  Training  in  Cross  Disciplinary Approaches to Non-Equilibrium Systems (CANES, Grant Nr.  EP/L015854/1)

\appendix

\section{Patterns mutual influence}
\label{app:mutual_interaction}
\begin{figure}
    \centering
    \subfloat[\label{fig_dots2}]{\includegraphics[scale=0.4]{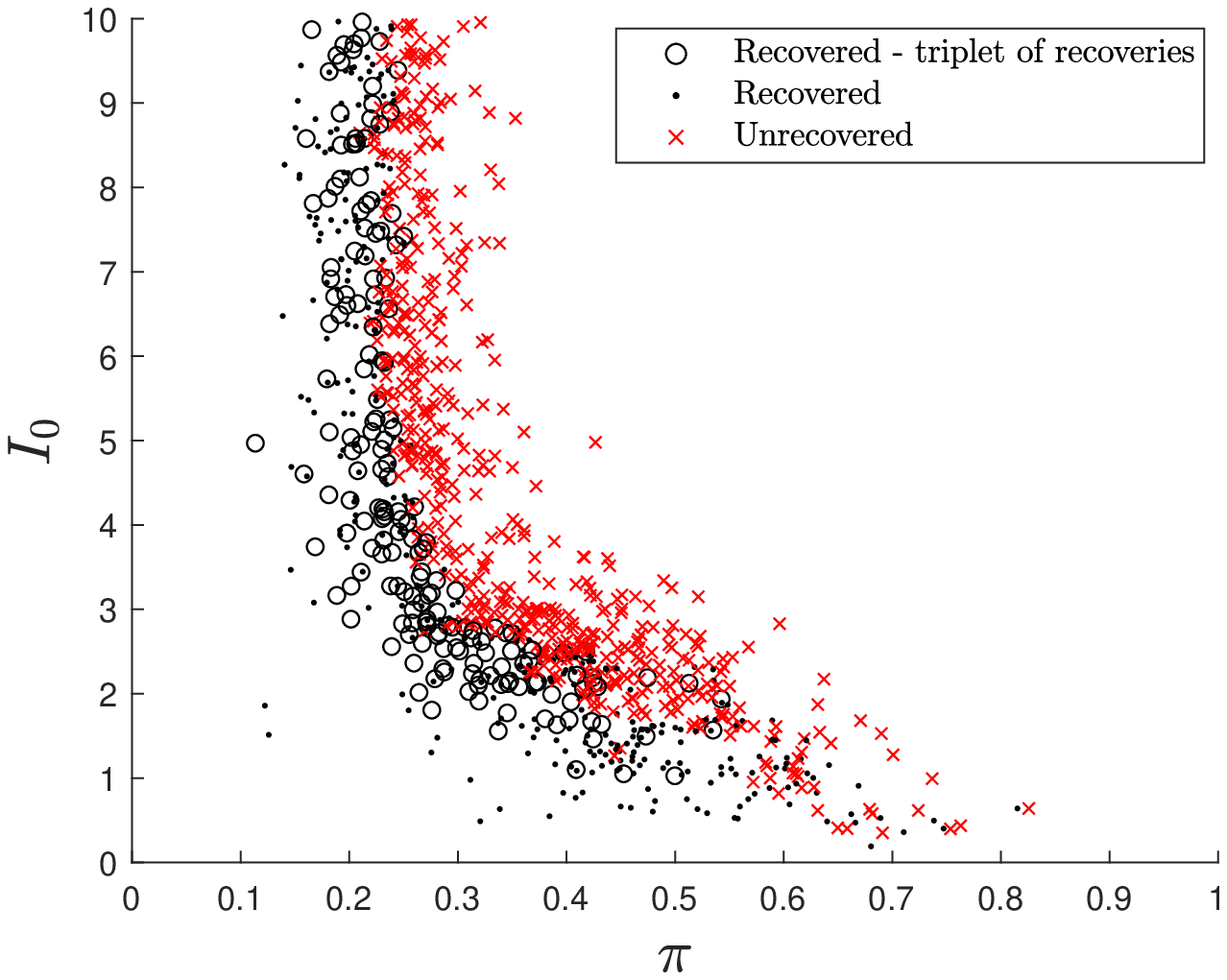}}
    \subfloat[\label{fig_dots3}]{\includegraphics[scale=0.4]{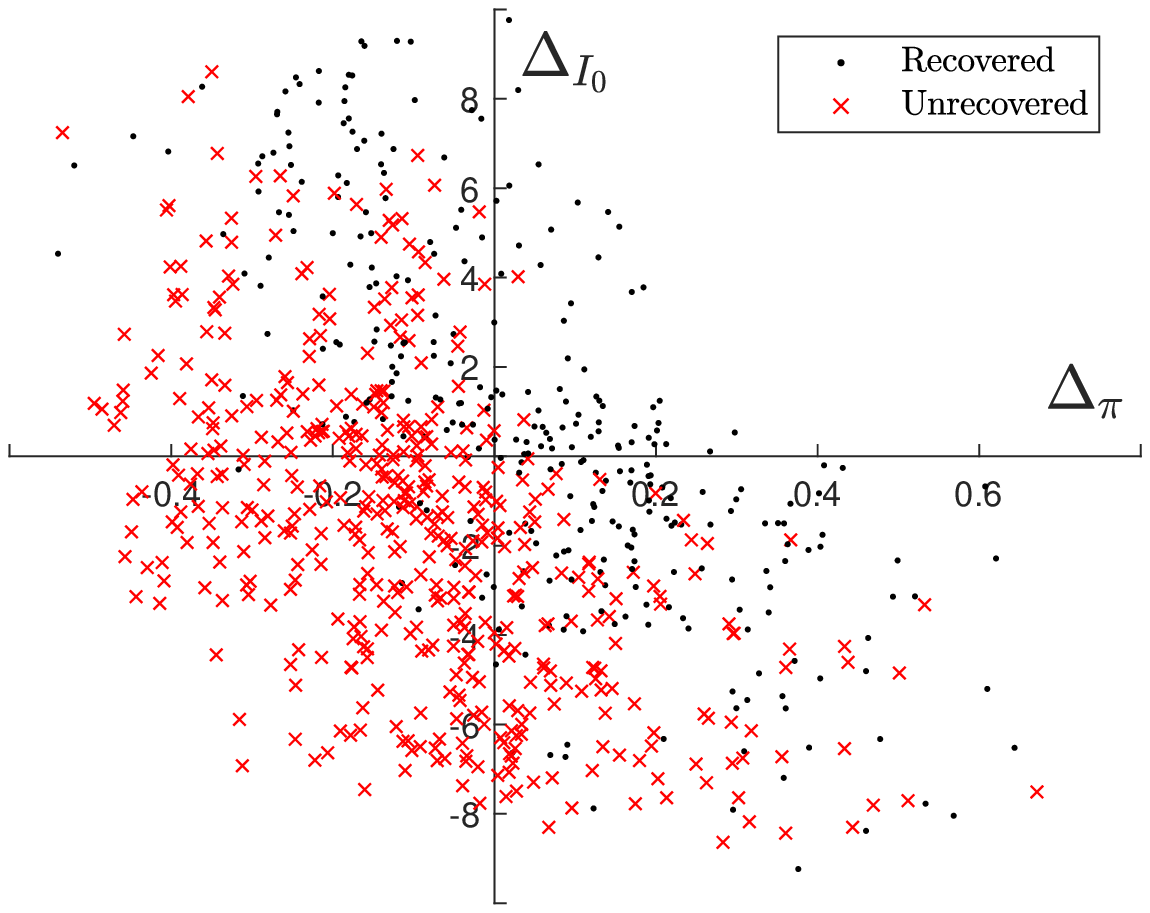}}\\
    \subfloat[\label{fig_dots4}]{\includegraphics[scale=0.4]{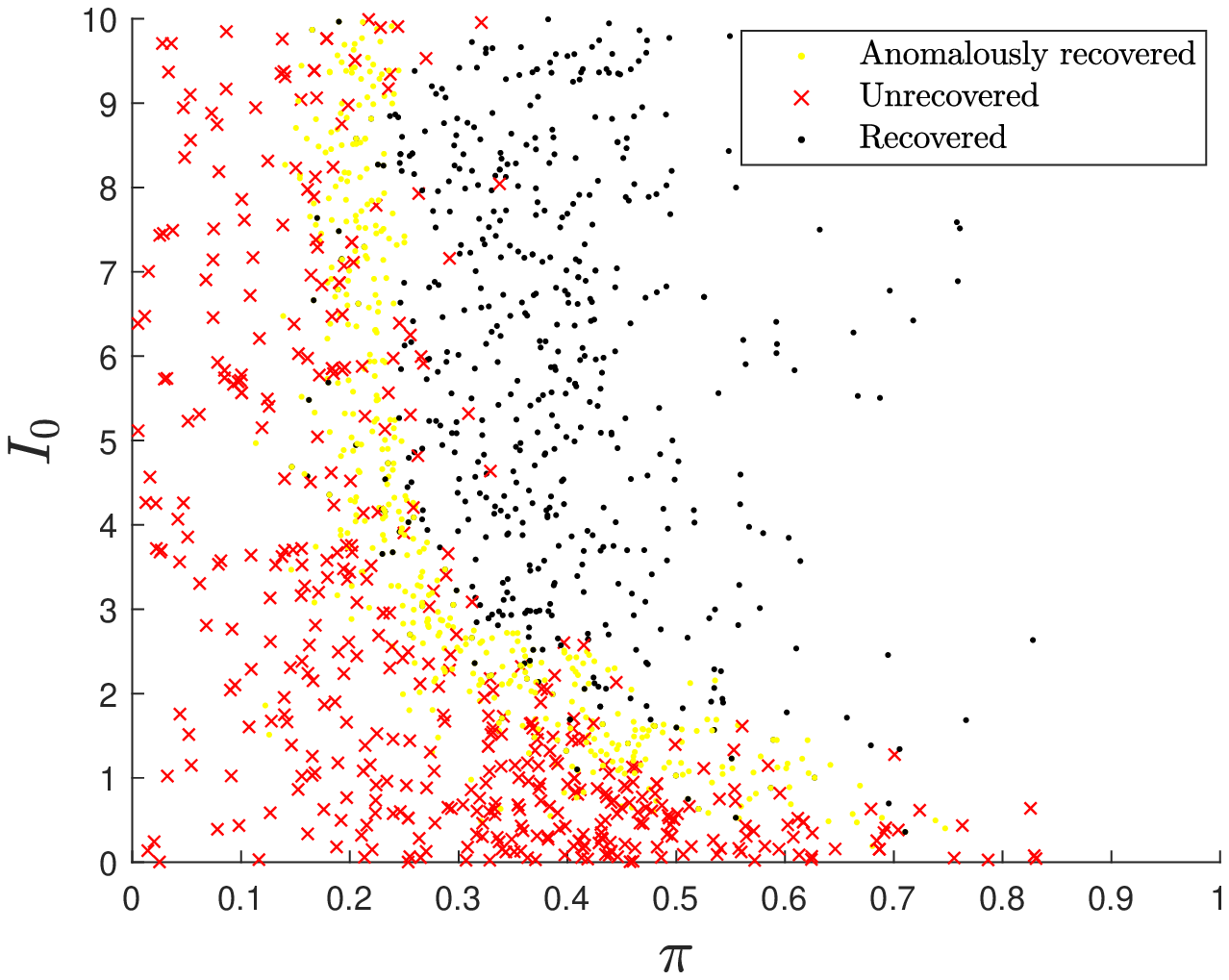}}
    \subfloat[\label{fig_dots5}]{\includegraphics[scale=0.4]{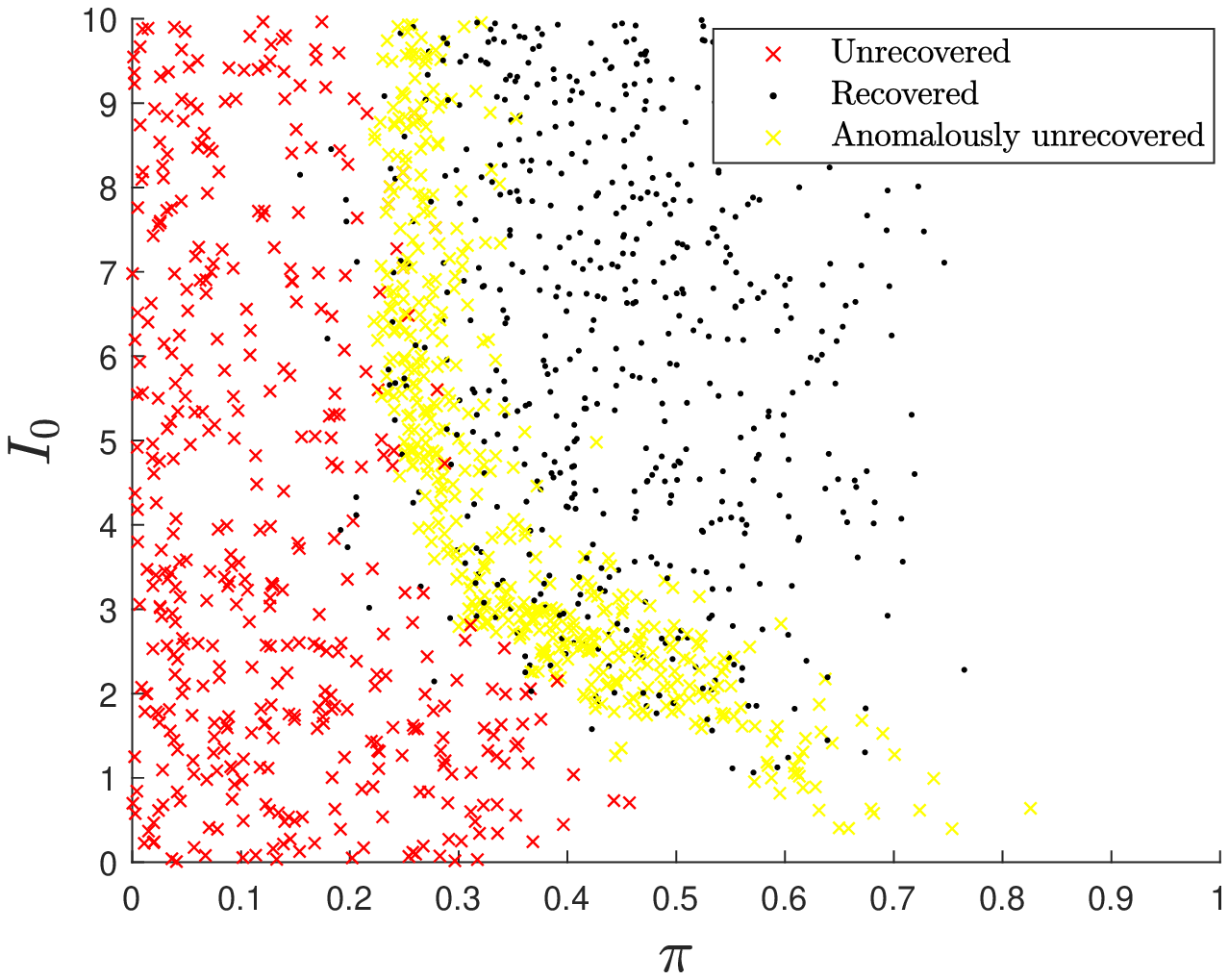}}
    \caption{(a) A selection of points from Fig. \ref{fig_dots1}  which have more than 50\% of neighbours of the opposite colour within a circle of radius of 0.05, plotted as a function of their signal strength and probability of presentation. Black markers correspond to recovered patterns and red markers to unrecovered patterns. Circles highlight patterns belonging to dynamics in which all three patterns presented have been recovered. Panel (b) shows a subset of the data of (a), excluding those that belong to triplets of patterns that are {\em all\/} recovered. Points are plotted as a function of their average distance from the patterns in the same run of the dynamics which have a different recovery state. (c) Yellow dots represent the recovered patterns in panel (a) while the red crosses and black dots represent respectively the unrecovered and recovered patterns in a run of the dynamics that gives rise to the yellow points. (d) Same as (c) but this time yellow crosses indicate unrecovered patterns from panel (a). The parameters of the simulation are $N= 100$, $J_0=8$ and $\Delta_0=17$.}
    \label{fig:random_triplets}
\end{figure}

To inspect the effect of mutual interactions between patterns of news concomitantly presented to the society, we studied several societal dynamics where random news are presented in triplets with 
random probability of appearance and random intensity.
In Fig. \ref{fig_dots1} we report the information on the final recovery or non-recovery of the patterns from the society, as a function of the probability and the intensity with which they were presented. In the graph we can notice a smooth transition region between a recovery zone and a non-recovery zone with some patterns within the recovery zone remaining un-recovered and, conversely, some patterns in the non-recovery zone apparently being recovered. Apart from an expected finite size effect, a factor which can also contribute to this phenomenon is the influence of the relative strengths and the relative probabilities of patterns presented during the same run of the dynamics. In fact, in Sec. \ref{sec:fixed_cond}, Fig. \ref{subfig_f2} we saw that, when there is one pattern in a triplet of patterns with a stronger amplitude, the other two are recovered less frequently than the strong pattern in question. In order to understand if this is a reason for anomalously recovered or un-recovered patterns, we propose to isolate outliers in the recovery and non-recovery zone which we identify as those patterns that have an anomalous behaviour when compared to their neighbours in the parameter space of frequency of occurrence and intensity, {\it i.e.} they are points corresponding to recovered patterns which are predominantly surrounded by those corresponding to un-recovered patterns and vice versa. Figure \ref{fig_dots2} shows a selection of points $\mu$ which have at least $50\%$ of neighbours $\mu'$ with  opposite retrieval outcome (and opposite colour in our representation) within distance $d=\sqrt{(I_0^\mu-I_0^{\mu'})^2+((\pi_\mu-\pi_{\mu'})\cdot 10)^2}=0.2$.
We multiplied the difference in probability to a factor 10 in order to account for the difference in scale between the probability which is randomly drawn between 0 and 1, and the signal strength which is randomly drawn between 0 and 10.
 
 To investigate whether mutual interactions between patterns influences recovery, we evaluate where the anomalous patterns are located relative to the other patterns in within their triplet which have the opposite recovery state. This means for example, that given a pattern that is unexpectedly recovered, we want to know how it is located relative to the un-recovered patterns presented within the same run of the dynamics. We thus calculate the average coordinate difference between the anomalously recovered point and the patterns in their triplets with opposite behaviour. 
 Given a recovered pattern $\mu=1$, assuming that both the patterns $\mu=2$ and $\mu=3$ are un-recovered, we calculate the coordinate difference in the $I_0$ direction as $\Delta_{I_0} = \frac{I_0^1-I_0^2}{2}+\frac{I_0^1-I_0^3}{2}$. If only $\mu=2$ is un-recovered then $\Delta_{I_0} = I_0^1-I_0^2$. Analogous formulae apply to the difference in the $\pi$ direction. 
We thus replot the points in Figure \ref{fig_dots2} as a function of the coordinate differences \footnote{Given that we are looking at the relative location of patterns behaving differently in the same triplet, we do not plot the points in dynamics where all the three patterns are recovered (little circles in Fig. \ref{fig_dots2}). Dynamics in which all the three patterns are not recovered are not present in the simulations performed.}.

Interestingly, we notice that recovered (unrecovered) patterns typically fall into their `natural' or expected top-right (bottom-left) regions when plotted in terms of their new coordinates.
In other words, unexpected recovery (or non-recovery) of patterns, as highlighted in the right panel, can be explained by the fact that they were appearing together with particularly weak (strong) or less (more) frequent patterns, which affected their likelihood of retrieval despite their own features. 

Additional evidence about the mutual influence of news retrievals can be obtained by looking directly at the statistics of news, which belong to the triplets of patterns that are anomalously retrieved or unretrieved by the society.
Figures \ref{fig_dots4} and \ref{fig_dots5} show in yellow respectively the anomalously recovered and unrecovered patterns of Fig. \ref{fig_dots2}, accompanied only by the patterns presented in the same run of the dynamics. The accompanying patterns are presented in red when they are not recovered and in black when they are recovered. 
In Fig.\ref{fig_dots4} we observe that anomalously recovered patterns are often accompanied by other recovered patterns with medium probability of appearance, which rarely reaches $\pi>0.5$.
Conversely, anomalously unrecovered patterns are accompanied by other recovered patterns with probability of appearance often in the range $0.5-0.8$, as visible in  Fig.\ref{fig_dots5}. These observations give  additional evidence to the hypothesis that very frequent news can hinder the ability of other news to be retrieved.
Moreover, we note that anomalously recovered patterns often appear concomitantly with very weak news with medium probability of appearance, see the dense cluster of unrecovered patterns with $\pi$ around $0.3-0.6$ and $I_0$ between $0$ and $1$ in Fig. \ref{fig_dots4}.
The same region of parameters is totally empty in Fig. \ref{fig_dots5}, which contain information about triplets of anomalously unretrieved patterns. This last observation supports the idea that the retrieval of a pattern is enhanced when accompanying patterns, which appear often in the dynamics, are weak enough.

Unfortunately, even in the simple case of societal dynamics involving only three patterns discussed here, these phenomena do not establish quantitative causal relations on the possibility to retrieve or not retrieve a pattern, although it becomes clear that fluctuations of retrieval probabilities cannot be simply ascribed to finite-size effects. 
We have, however, collected indirect evidence that the relative strength and probability of the patterns in the system can influence the likelihood of their recovery. In particular, a pattern which has a probability and/or a strength which is small in magnitude can be recovered if the other patterns in the same dynamics are weaker in probability or signal strength. Similarly, patterns in a triplet of patterns with medium/high probability and/or high signal strength can end up being unexpectedly unrecovered.

\section{Replica calculations for an infinite flow of patterns}
\label{app:replica}
In this Appendix we derive the system of equations \eqref{fpes}, which defines the order parameters of the model in the noiseless limit of Eq.\,\eqref{maineq}. The society described by this model is subject to an infinite sequence of external signals representing exposure to an infinite history of news items (Eq.\,\eqref{signal-infinite}), as introduced in Sec. \ref{sec:infinite_num_patt}. Using replica calculations, we obtain the free-energy of the model and we derive and solve the corresponding fixed-point equations in the zero-temperature limit. The calculation follows standard reasoning. For further details, we refer to \cite{kuhn1993statistical}.

\subsection{Replica Approach}
We start assuming that the asymptotic state of our system is macroscopically correlated with at most one of the presented patterns, taken to be pattern $\mu$. We want to obtain the average of the free energy 
of the system taken over the randomness in all but pattern $\mu =\alpha N$. 
This is done using replica trick, for which we can write the free energy in the form
\be
 f(\beta) = - \frac{1}{\beta} \lim_{N\to\infty}\lim_{n\to 0}~ (Nn)^{-1} \ln 
\overline{Z_N^n}
\ee
where $\beta$ is the reciprocal of the thermodynamic temperature which tends to infinity in the noiseless limit, while $n$ is the number of replica and
\be
Z_N^n = \int\prod_{i\sigma} dv_{i\sigma} \exp\Big\{ 
-\beta \sum_{\sigma=1}^n \cH(\{\bm v_\sigma\}) \Big\}\ .
\ee
is the replicated partition function. One can write the Hamiltonian of a replica $\sigma$ appearing in this expression as follows
\bea
\cH(\{\bm v_\sigma\}) &=& -\frac{N J_0 \tilde{\gamma}_0}{2}  
{m^\mu_{\sigma}}^2\, \re^{-\tilde{\gamma}_0 \alpha} 
-  \frac{J_0 \tilde{\gamma}_0}{2}\sum_{\nu(\neq \mu)} {X^\nu_{\sigma}}^2 \,
 \re^{-\tilde{\gamma}_0 \nu/N}  \nn\\ 
& & +\frac{J_0}{2} \bigg(1-\re^{-\tilde{\gamma}_0\alpha_{\rm max}} \bigg)\sum_{i}v_{i\sigma}^2 
-\sum_{i}  I_i v_{i\sigma} + 
\sum_{i}  G(v_{i\sigma})\ .
\eea
Here
\bea
m^\mu_{\sigma}&=&\frac{1}{N} \sum_i \xi_i^\mu v_{i\sigma} \nn\\
X^\nu_{\sigma}&=&\frac{1}{\sqrt{N}} \sum_i \xi_i^\nu v_{i\sigma}
\quad , \qquad \nu (\neq \mu)\ ,\nn
\eea
and $G(v)$ is the integrated inverse input-output relation
\be
G(v) = \int^v {\rm d} v' g^{-1}(v')\ .
\ee
The disorder due to the patterns $\nu (\neq \mu)$ appears only through the $X^\nu_{\sigma}$, which for any fixed configuration $\{\bm v_\sigma\}$ are Gaussian random
variables of zero mean and covariance
\be
\langle X^\nu_{\sigma} X^{\nu'}_{\sigma'}\rangle =  
\frac{\delta_{\nu\nu'}}{N}\sum_i v_{i\sigma} v_{i\sigma'} = 
\delta_{\nu\nu'} q_{\sigma\sigma'}\ .
\ee
The average over the patterns disorder can therefore be performed as a Gaussian integral, 
resulting in
\bea
\overline{Z_N^n} &=& \int \prod_{i\sigma} dv_{i\sigma}
\exp\Bigg\{N\Bigg[\beta \frac{J_0 \tilde{\gamma}_0}{2} \sum_\sigma 
{m^\mu_{\sigma}}^2\, \re^{-\tilde{\gamma}_0 \alpha}
 +\\
 &\ &-\frac{1}{2}\int_0^{\alpha_{\rm max}}\rd x\, {\rm tr} \ln\Big (\bbbone - \beta  J_0 \tilde{\gamma}_0\, 
\re^{-\tilde{\gamma}_0 x}Q\Big)\Bigg]\nn \\
& &-\frac{\beta J_0}{2} \bigg(1- \re^{-\tilde{\gamma}_0\alpha_{\rm max}} \bigg)\sum_{i,\sigma}v_{i\sigma}^2 
+\beta \sum_{i,\sigma} I_i v_{i\sigma} -\beta \sum_{i\sigma}  G(v_{i\sigma})
\Bigg\} \nn
\eea
in which $Q$ is a matrix with elements $q_{\sigma\sigma'}$.
The standard procedure now is to introduce the overlaps $m^\mu_\sigma$ and 
Edwards-Anderson order parameters $q_{\sigma\sigma'}$ as integration variables,
using conjugate variables for Fourier representations of $\delta$-functions that enforce
the definition of these order parameters. This results in
\bea
\langle Z_N^n\rangle  
&=&\int \prod_{\sigma} \frac{\rd{m^\mu_{\sigma}} \rd\hat{m^\mu_{\sigma}}}{2\pi/N} \prod_{\sigma\sigma'}
\frac{\rd q_{\sigma\sigma'} \rd\hat q_{\sigma\sigma'}}{2\pi/N} 
\exp\Bigg\{N\Bigg[\beta\frac{ J_0\tilde{\gamma}_0}{2} \sum_\sigma 
{m^\mu_{\sigma}}^2\, \re^{-\tilde{\gamma}_0 \alpha}  \nn\\
& & - \frac{1}{2}\int_0^{\alpha_{\rm max}}\rd x\, {\rm tr} \ln\Big (\bbbone - \beta  J_0 \tilde{\gamma}_0\, 
\re^{-\tilde{\gamma}_0 x}Q\Big) \nn \\
&\ &-\sum_\sigma \ri\hat m_\sigma m_\sigma - \sum_{\sigma\sigma'} \ri \hat 
q_{\sigma\sigma'} q_{\sigma\sigma'} \nn\\
& & + \frac{1}{N} \sum_i \ln \int \prod_{\sigma} d v_{\sigma} \exp 
\Bigg\{\sum_\sigma \ri\hat m_\sigma \xi_i^\mu v_\sigma 
+\sum_{\sigma\sigma'} \ri\hat q_{\sigma\sigma'} v_\sigma v_{\sigma'} 
\nn \\
& & -\frac{\beta J_0}{2} \bigg(1-\re^{-\tilde{\gamma}_0\alpha_{\rm max}} \bigg)\sum_{\sigma}v_{\sigma}^2 
 +\beta  I_i \sum_{\sigma}  v_{\sigma} - \beta  \sum_{\sigma} G(v_{\sigma})\Bigg\}\Bigg]
\Bigg\}\ .
\label{finalZn}
\eea
This is now of a form that can be evaluated by the saddle-point method.
In the thermodynamic limit, the (empirical) average over single site free
energies $\frac{1}{N} \sum_i \ln \int \prod_{\sigma} d v_{\sigma} 
\exp\{\dots\}$ appearing in the last two lines of (\ref{finalZn}) will, by the law of large numbers,
converge to a joint average over all forms of on-site randomness present in
the average, i.e., the $I_i$, and the $\xi_i^\mu$. 

The stationarity requirement on the exponent w.r.t. variations of the conjugate variables gives
two self-consistency equations
\be
{m^\mu_{\sigma}} =\ldd \xi^\mu\, \langle v_\sigma \rangle\ \rdd
\quad , \qquad 
q_{\sigma\sigma'} = \ldd\ \langle v_\sigma v_{\sigma'} \rangle\ \rdd \ ,
\label{fpe}
\ee
while the stationarity requirement w.r.t. the order parameters $m^\mu_{\sigma}$ and 
$q_{\sigma\sigma'}$ results in
\bea
\ri \hat{m^\mu_{\sigma}}& =& \beta J_0 \tilde{\gamma}_0{m^\mu_{\sigma}}\, \re^{-\tilde{\gamma}_0 \alpha}\  ,
\label{fpeconjm}\\ 
\ri \hat q_{\sigma\sigma'} &=&  \frac{\beta J_0 \tilde{\gamma}_0}{2}\int_0^{\alpha_{\rm max}}\rd x\, 
 \re^{-\tilde{\gamma}_0 x}\,
\Big (\bbbone - \beta  J_0 \tilde{\gamma}_0\, \re^{-\tilde{\gamma}_0 x}Q\Big)^{-1}_{\sigma\sigma'}\ .
\label{fpeconjq}
\eea
Inner averages in (\ref{fpe}) denote thermal averages over effective 
$i$-dependent replicated single-site Hamiltonians defined via (\ref{finalZn});
after inserting the values of the conjugate order parameters as defined via Eqs.
(\ref{fpeconjm}) and (\ref{fpeconjq}) , it is seen to take the form
\bea
H^{(i)}_{\rm eff}(\{v_\sigma\})&=&-J_0 \tilde{\gamma}_0 \, \re^{-\tilde{\gamma}_0 \alpha} \sum_\sigma \xi_i^\mu{m^\mu_{\sigma}}  v_\sigma\nn\\
&& -  \frac{J_0 \tilde{\gamma}_0}{2} \sum_{\sigma\sigma'} \int_0^{\alpha_{\rm max}}\rd x\, 
 \re^{-\tilde{\gamma}_0 x}\,
\Big (\bbbone - \beta  J_0 \tilde{\gamma}_0\, \re^{-\tilde{\gamma}_0 x}Q\Big)^{-1}_{\sigma\sigma'}
 v_\sigma v_{\sigma'}  \nn\\
&& + \frac{ J_0}{2} \bigg(1- \re^{-\tilde{\gamma}_0\alpha_{\rm max}} \bigg)\sum_{\sigma}v_{\sigma}^2
 - I_i \sum_{\sigma} v_{\sigma} + \sum_{\sigma} G(v_{\sigma})
\label{Hieff}
\eea
The $i$ dependence originates from the on-site disorder in (\ref{finalZn}),
and the outer average denoted by double angle brackets in (\ref{fpe})
denotes an average over the joint distribution of this remaining on-site 
disorder.

\subsection{Replica Symmetry}
Assuming replica symmetry (RS) for the solutions of the fixed point equations, with
\be
{m^\mu_{\sigma}} = m\quad ,\quad q_{\sigma\sigma} = q_d \quad ,\quad {\rm and}~~~~ 
q_{\sigma\sigma'} = q \quad {\rm for}~~~ \sigma\ne \sigma' \ ,
\label{RS}
\ee
one can decouple the replica and take the $n\to 0$-limit of the theory as
required. 
Using the shorthand $a(x) = \beta  J_0 \tilde{\gamma}_0\, \re^{-\tilde{\gamma}_0 x}$, we notice that the term
coupling replica in the effective single-site Haminltonian $H^{(i)}_{\rm eff}(\{v_\sigma\}) $ in Eq.\,\eqref{Hieff} is
\begin{eqnarray}
 \frac{1}{2}\sum_{\sigma\sigma'}&&\hspace{-0.5cm} \int_0^{\alpha_{\rm max}}\hspace{-5mm}\rd x\, a(x)
\Big (\bbbone - a(x) Q\Big)^{-1}_{\sigma\sigma'}  v_\sigma v_{\sigma'} \\
&&= \frac{1}{2} \sum_{\sigma\sigma'} \int_0^{\alpha_{\rm max}}\hspace{-5mm}\rd x\, a(x)
\Big (\big(1- a(x)(q_d-q)\big)\bbbone - a(x) q\fbbone\Big)^{-1}_{\sigma\sigma'}  v_\sigma v_{\sigma'} \nn 
\end{eqnarray}
in which $\bbbone$ is the $n\times n$ unit matrix, while $\fbbone$ is the $n\times n$ with all elements
equal to 1. Using the algebra of these matrices one finds
$$
(A \bbbone - B\fbbone)^{-1} = \frac{1}{A} \bbbone + \frac{B}{A(A+Bn)} \fbbone \quad\longrightarrow \quad
 \frac{1}{A} \bbbone+ \frac{B}{A^2} \fbbone \ , \qquad \mbox{as}\ n\to 0\ .
$$ 
We thus get
\bea
 \frac{1}{2} &&\hspace{-1cm}\sum_{\sigma\sigma'}\int_0^{\alpha_{\rm max}}\hspace{-5mm}\rd x\, a(x)
\Big (\big(1- a(x)(q_d-q)\big)\bbbone - a(x) q\fbbone\Big)^{-1}_{\sigma\sigma'}  v_\sigma v_{\sigma'}\nonumber \\
&=&\frac{1}{2} \int_0^{\alpha_{\rm max}}\hspace{-5mm}\rd x\, \frac{a(x)}{1- a(x)(q_d-q)} \  \sum_{\sigma} v_\sigma^2\nn\\
 & & + \frac{1}{2} \int_0^{\alpha_{\rm max}}\hspace{-5mm}\rd x\, \frac{a(x)^2 q}{\big(1- a(x)(q_d-q)\big)^2} 
\Big(\sum_{\sigma} v_\sigma\Big)^2
\label{RS1}
\eea
Exploiting the fact that $a'(x)=-\tilde{\gamma}_0 a(x)$, we find (assuming $\alpha_{\rm max}\to \infty$) the first integral to give
\be
\frac{1}{2} \int_0^{\infty}\rd x\, \frac{a(x)}{1- a(x)(q_d-q)} 
= - \frac{1}{2}\,\frac{\ln \big(1- \beta J_0 \tilde{\gamma}_0 (q_d-q)\big)}{\tilde{\gamma}_0 (q_d-q)}\  
\ee
To evaluate the second integral, use
\be
\frac{\rd}{\rd x} \Bigg[\frac{a(x)}{1-a(x)(q_d-q)}\Bigg] = \frac{-\tilde{\gamma}_0 a(x)}{1-a(x)(q_d-q)} - \frac{a(x)^2 \tilde{\gamma}_0 (q_d-q)}{\big(1- a(x)(q_d-q)\big)^2} \ ,
\ee
hence
\bea
\frac{1}{2} &&\hspace{-0.8cm}\int_0^{\infty}\!\!\rd x\, \frac{a(x)^2 q}{\big(1- a(x)(q_d-q)\big)^2} \nn \\
&=& \frac{1}{2} \frac{q}{\tilde{\gamma}_0 (q_d-q)}\Bigg[\frac{\beta J_0 \tilde{\gamma}_0}
{1-\beta J_0 \tilde{\gamma}_0 (q_d-q)} +  \frac{\ln \big(1- \beta J_0 \tilde{\gamma}_0 (q_d-q)\big)}{(q_d-q)}\Bigg]
\nn\\
&=& \frac{1}{2} \frac{\beta J_0 q}{q_d-q} \Bigg[\frac{1}
{\big(1-\beta J_0 \tilde{\gamma}_0 (q_d-q)\big)} +  
\frac{\ln \big(1- \beta J_0 \tilde{\gamma}_0 (q_d-q)\big)}{\beta J_0 \tilde{\gamma}_0 (q_d-q)}\Bigg]
\eea
Although it is not obvious at this point, this integral is positive by construction.

It is expected that $C=\beta(q_d-q)$  will remain finite in the $\beta\to\infty$-limit we are interested in.

This gives
$$
\frac{1}{2} \int_0^{\infty}\rd x\, \frac{a(x)}{1- a(x)(q_d-q)} 
= - \frac{\beta J_0}{2}\,\frac{\ln \big(1- J_0\tilde{\gamma}_0 C\big)}{J_0\tilde{\gamma}_0 C}\  
$$
and
\bea
\frac{1}{2} \int_0^{\infty}\!\!\rd x\, \frac{a(x)^2 q}{\big(1- a(x)(q_d-q)\big)^2} 
&=&\frac{1}{2} \frac{(\beta J_0)^2  q}{J_0 C} \Bigg[\frac{1}{1-J_0\tilde{\gamma}_0 C} + \frac{\ln \big(1- J_0\tilde{\gamma}_0 C\big)}{J_0\tilde{\gamma}_0 C}\Bigg] \nn \\
&\equiv& \frac{1}{2} (\beta J_0)^2 r
\label{defr}
\eea
The coupling between replica through a complete square embodied in the second contribution in \eqref{RS1}
is then dealt with in the usual way by Gaussian linearisation.
 
The fixed point equations for $m$, $q$ and $C$ finally take the form
\be
m = \ldd \xi^\mu~ \langle v \rangle \rdd \quad , \quad
C = \frac{1}{J_0 \sqrt{r}} \ldd z~\langle v \rangle \rdd \quad , \quad
q = \ldd \langle v \rangle^2 \rdd \ ,
\label{fpeRS}
\ee
in which $r$ is defined through Eq.\,\eqref{defr}:
\begin{equation}
r = \frac{ q}{J_0 C} \Bigg[\frac{1}{1-J_0\tilde{\gamma}_0 C} + \frac{\ln \big(1- J_0\tilde{\gamma}_0 C\big)}{J_0\tilde{\gamma}_0 C}\Bigg]\ .
\end{equation}
The inner average in Eq.s \ref{fpeRS} is now a thermal average over 
effective  RS single-site Hamiltonians of the form
\be
H_{RS}=- \Big(\xi^\mu m \tilde{J_0}e^{-\gamma_0 \Delta_0 \alpha} + J_0\sqrt{r}~z + I\Big)~v
+ \frac{J_0}{2}~\Big(1 + \frac{\ln(1-J_0\tilde{\gamma}_0 C)}{J_0\tilde{\gamma}_0 C}\Big)~v^2  + G(v)
\label{HRS}
\ee
while outer averages are over the on-site disorder {\em and\/} over the 
additional zero-mean unit-variance Gaussian $z$ appearing in $H_{\rm RS}$. In the limit for $\beta \to \infty$, given a generic function $F$ the inner averages can be written as:
\be
\langle F(v)\rangle = F(\hat{v})
\ee
where $\hat{v}$ is the value of $v$ which minimizes $H_{RS}$
\bea\label{v_hat}
\hat v(\xi^\mu,I,z) = g\bigg(m \xi^\mu J_0\tilde{\gamma}_0e^{-\tilde{\gamma}_0 \alpha} + J_0 \sqrt{r}z + I -  J_0 \Big(1 + \frac{\ln(1-J_0\tilde{\gamma}_0 C)}{J_0\tilde{\gamma}_0 C}\Big) \hat v\bigg ) 
\ .\label{hvz}
\eea
We are thus left with the problem of solving a $z$-dependent fixed-point equation, Eq.\,\eqref{hvz}, {\em within\/} the system \eqref{fpeRS} of fixed-point equations,  which is avoided by transforming the Gaussian $z$-distribution into a $\hat v$-distribution and then taking the averages respect to $\hat{v}$. For a vanishing signal, i.e. $I=0$, the resulting $p(\hat{v})$ will be given by
\begin{eqnarray}\label{p(v)}
      p(\hat{v})&=&\frac{\e^{-z^2/2}}{\sqrt{2\pi}}\frac{\mathrm{d} z}{\mathrm{d}\hat{v}}\ , \nonumber \\
      &=& \frac{\e^{-z^2/2}}{J_0\sqrt{2\pi r}}\left[\frac{\sqrt{\pi}}{2}\e ^{{\erf^{-1}(\hat{v})}^2}  + J_0\left(1+\frac{\ln(1-J_0\tilde{\gamma}_0C)}{J_0\tilde{\gamma}_0C}\right)\right]\ ,
\end{eqnarray}
with:
\begin{equation}\label{z(v)}
    z=\frac{1}{J_0\sqrt{r}}\left[\erf^{-1}(\hat{v})-m\xi^\mu J_0\tilde{\gamma}_0\e^{-\tilde{\gamma}_0\alpha}-I+J_0\hat{v}\left(1+\frac{\ln(1-J_0\tilde{\gamma}_0C)}{J_0\tilde{\gamma}_0C}\right) \right]\ .
\end{equation}
If instead we consider a Gaussian signal, as described in Eq.\,\eqref{distorted_signal}, the value of $\hat{v}$ can be written as:

\be
\hat v(\xi^\mu,z,z') = g\bigg(\xi^\mu (m J_0\tilde{\gamma}_0e^{-\tilde{\gamma}_0 \alpha} + \widetilde{I_0}) + J_0 \sqrt{r}z + \sigma_I z' - J_0 \Big(1 + \frac{\ln(1-J_0\tilde{\gamma}_0 C)}{J_0\tilde{\gamma}_0 C}\Big) \hat v\bigg ) 
\ .
\ee
in which $z, z'$ are two independent standard Gaussians. Using the fact that the sum of Gaussians is itself Gaussian (with variances given by the sum of variances), we can write
\be
\hat v(\xi^\mu,z) = g\bigg(\xi^\mu (m J_0\tilde{\gamma}_0e^{-\tilde{\gamma}_0 \alpha} + \widetilde{I_0}) +  \sqrt{J_0^2 r+ \sigma_I^2}\, z  -  J_0 \Big(1 + \frac{\ln(1-J_0\tilde{\gamma}_0 C)}{J_0\tilde{\gamma}_0 C}\Big) \hat v\bigg ) 
\ .
\ee
so
\begin{equation}
z=\frac{1}{\sqrt{J_0^2r+\sigma_I^2}}\left[\erf^{-1}(\hat{v})-\xi^\mu (m J_0\tilde{\gamma}_0\e^{-\tilde{\gamma}_0\alpha}+\widetilde{I_0})+J_0\hat{v}\left(1+\frac{\ln(1-J_0\tilde{\gamma}_0C)}{J_0\tilde{\gamma}_0C}\right) \right]\ .
\end{equation}
and 
\begin{equation}
      p(\hat{v}) = \frac{\e^{-z^2/2}}{\sqrt{2\pi} \sqrt{J_0^2 r +\sigma_I^2}}\left[\frac{\sqrt{\pi}}{2}\e ^{{\erf^{-1}(\hat{v})}^2}  + J_0\left(1+\frac{\ln(1-J_0\tilde{\gamma}_0C)}{J_0\tilde{\gamma}_0C}\right)\right]\ .
\end{equation}

\section{Methods}
\label{app:methods}
\subsection{Simulations}
In this paper we simulate the dynamics of the society described by Eq.\,\eqref{maineq} in order to measure the spontaneous retrieval of opinion patterns induced by the presentation of different kind of external signals.

In order to do this, we discretize the Eq.(\ref{maineq}) with a time step $\D t=0.1$, where smaller $\D t$ were seen not to significantly change the results. At each time step we use Euler method to calculate the preference of each agent $u_i$, and the couplings $J_{ij}$, discretizing with the same time step its differential equation (obtained differentiating Eq.(\ref{eqJs})):
\begin{equation}\label{dynrule}
\dot J_{ij}(t)= \gamma \left[\frac{J_0}{N}v_i(t)v_j(t) - J_{ij}(t)\right]\ .
\end{equation}
Even if the Euler method is a simple integration method, it is accurate enough for our purposes.

There are many parameters to be set up for these simulations.
In Sec. \ref{sec:random} we simulate a society which receives $p=3$ patterns in a random order. 
The size of the system is set to be $N=100$, the strength of the interactions $J_0=6$ and the memory factor $\gamma = 10^{-3}$. Other parameters, as the time length of the external signal and the amplitude $I_0$ of the polarizing signal, change in different simulations and are indicated under the corresponding figures. The number of agents $N=100$ is a good approximation of the thermodynamic limit at finite $p$. Although these numbers seem small, they produce results that are representative of the $N\to \infty$ limit with $p\ll N$.

In Sec. \ref{sec:infinite_num_patt} we perform the same kind of simulations to describe a society which receives an infinite number of external news. 
In this case $J_0$ is fixed to 0.2, a value at which was possible to obtain analytical results to be compared to the simulations. The signal strength was chosen to be $I_0 = 10$ while the size of the system explored where $N=200$ and $N=800$. The value of $\tilde{\gamma}_0$ is chosen to be 15, close to the point at which the corresponding critical capacity is maximum. The value of $\tilde{\gamma}$ is varied through different simulations, with only the value of $\Delta_0$ changing and $\gamma_0$ always kept at 1.
Through all the paper we used a low noise level, with variance  $\sigma^2=0.01$, to ensure that non-trivial collective states can emerge.
All simulations start with random initial conditions $u_i \sim \mathcal{N}(0, \sigma^2/2)$ which would be the equilibrium distribution in a non-interacting system without external signal.

\subsection{Numerical solutions}
Numerical solutions of Eq.\,\eqref{fpes} are obtained iteratively. The double angle bracket are evaluated as an average over $\xi^\mu = \pm 1$ and an integral over $p(\hat{v})$ (Eq.\,\eqref{p(v)}).

We would like to find a solution close to $m = 1$ so we start with an initial guess of $m \sim 1$, $c\sim0$ and $q\sim 1$, and we iterate as follow:
\bea
m(i+1) &=& \rho m(i) + (1-\rho)\frac{1}{2}\left[\int_{-1}^{1}\mbox{d} \hat{v} \hat{v} p_{+}(\hat{v})- \int_{-1}^{1}\mbox{d} \hat{v}\hat{v} p_{-}(\hat{v})\right]\nonumber\ ,\\
C(i+1) &=& \rho C(i) + (1-\rho)\frac{1}{2J_0\sqrt{r}}\left[\int_{-1}^{1}\mbox{d} \hat{v} zp_{+}(\hat{v})+ \int_{-1}^{1}\mbox{d} \hat{v}  zp_{-}(\hat{v})\right]\nonumber \ ,\\
q(i+1) &=& \rho q(i) + (1-\rho)\frac{1}{2}\left[\int_{-1}^{1}\mbox{d} \hat{v} \hat{v}^2 p_{+}(\hat{v})+ \int_{-1}^{1}\mbox{d} \hat{v} \hat{v}^2 p_{-}(\hat{v})\right]\ ,\label{iterative-solution}
\eea
where $z$ is expressed as a function of $\hat{v}$ as described by Eq.\,\eqref{z(v)}. The notation $p_{\pm}(\hat{v})$ indicates the distribution $p(\hat{v})$ evaluated at $\xi^\mu = \pm 1$. We consider the iterative algorithm converged when the difference between the values found at step $i+1$ and $i$ is smaller than $10^{-9}$. Numerical solutions of the set are found for different values of $\alpha$ and $\tilde{\gamma}_0$. For each couple of parameters, the iterative algorithm is initialized in the solution found in the closest point in the parameter space.

We should finally notice that for
\be \label{gap_condition}
J_0\left(1+\frac{\log(1-J_0\tilde{\gamma}_0C)}{J_0\tilde{\gamma}_0}\right) < 0
\ee 
the slope of the error function appearing in Eq.\,\eqref{v_hat} induces the solution of $\hat{v}$ to be discontinuous. In particular $\hat{v}$ is negative from -1 to $-v_0$, jumps to a positive value $v_0$ which must be found numerically, and remains positive from $v_0$ to $1$. In this case the integrals appearing in Eq.\,\eqref{iterative-solution} must be split and taken from $-1$ to $-v_0$ and from $v_0$ to 1. For the parameters used in this paper the condition in Eq.\,\eqref{gap_condition} is not met, so this split is not necessary.

\bibliography{mybib}

\begin{thebibliography}{43}
\expandafter\ifx\csname natexlab\endcsname\relax\def\natexlab#1{#1}\fi
\providecommand{\url}[1]{\texttt{#1}}
\providecommand{\href}[2]{#2}
\providecommand{\path}[1]{#1}
\providecommand{\DOIprefix}{doi:}
\providecommand{\ArXivprefix}{arXiv:}
\providecommand{\URLprefix}{URL: }
\providecommand{\Pubmedprefix}{pmid:}
\providecommand{\doi}[1]{\href{http://dx.doi.org/#1}{\path{#1}}}
\providecommand{\Pubmed}[1]{\href{pmid:#1}{\path{#1}}}
\providecommand{\bibinfo}[2]{#2}
\ifx\xfnm\relax \def\xfnm[#1]{\unskip,\space#1}\fi
\bibitem[{Castellano et~al.(2009)Castellano, Fortunato, and
  Loreto}]{castellano2009statistical}
\bibinfo{author}{C.~Castellano}, \bibinfo{author}{S.~Fortunato},
  \bibinfo{author}{V.~Loreto}, \bibinfo{journal}{Reviews of Modern Physics}
  \bibinfo{volume}{81} (\bibinfo{year}{2009}) \bibinfo{pages}{591}.
\bibitem[{Flache et~al.(2017)Flache, M{\"a}s, Feliciani, Chattoe-Brown,
  Deffuant, Huet, and Lorenz}]{flache2017models}
\bibinfo{author}{A.~Flache}, \bibinfo{author}{M.~M{\"a}s},
  \bibinfo{author}{T.~Feliciani}, \bibinfo{author}{E.~Chattoe-Brown},
  \bibinfo{author}{G.~Deffuant}, \bibinfo{author}{S.~Huet},
  \bibinfo{author}{J.~Lorenz}, \bibinfo{journal}{Journal of Artificial
  Societies and Social Simulation} \bibinfo{volume}{20} (\bibinfo{year}{2017}).
\bibitem[{Baronchelli(2018)}]{baronchelli2018emergence}
\bibinfo{author}{A.~Baronchelli}, \bibinfo{journal}{Royal Society open science}
  \bibinfo{volume}{5} (\bibinfo{year}{2018}) \bibinfo{pages}{172189}.
\bibitem[{Galam et~al.(1982)Galam, Gefen, and Shapir}]{galam1982sociophysics}
\bibinfo{author}{S.~Galam}, \bibinfo{author}{Y.~Gefen},
  \bibinfo{author}{Y.~Shapir}, \bibinfo{journal}{Journal of Mathematical
  Sociology} \bibinfo{volume}{9} (\bibinfo{year}{1982}) \bibinfo{pages}{1--13}.
\bibitem[{Galam and Moscovici(1991)}]{galam1991towards}
\bibinfo{author}{S.~Galam}, \bibinfo{author}{S.~Moscovici},
  \bibinfo{journal}{European Journal of Social Psychology} \bibinfo{volume}{21}
  (\bibinfo{year}{1991}) \bibinfo{pages}{49--74}.
\bibitem[{Borghesi and Bouchaud(2007)}]{borghesi2007songs}
\bibinfo{author}{C.~Borghesi}, \bibinfo{author}{J.-P. Bouchaud},
  \bibinfo{journal}{Quality \& Quantity} \bibinfo{volume}{41}
  (\bibinfo{year}{2007}) \bibinfo{pages}{557--568}.
\bibitem[{Clifford and Sudbury(1973)}]{clifford1973model}
\bibinfo{author}{P.~Clifford}, \bibinfo{author}{A.~Sudbury},
  \bibinfo{journal}{Biometrika} \bibinfo{volume}{60} (\bibinfo{year}{1973})
  \bibinfo{pages}{581--588}.
\bibitem[{Holley and Liggett(1975)}]{holley1975ergodic}
\bibinfo{author}{R.~A. Holley}, \bibinfo{author}{T.~M. Liggett},
  \bibinfo{journal}{The Annals of Probability}  (\bibinfo{year}{1975})
  \bibinfo{pages}{643--663}.
\bibitem[{Galam(2002)}]{galam2002minority}
\bibinfo{author}{S.~Galam}, \bibinfo{journal}{The European Physical Journal
  B-Condensed Matter and Complex Systems} \bibinfo{volume}{25}
  (\bibinfo{year}{2002}) \bibinfo{pages}{403--406}.
\bibitem[{Martins and Galam(2013)}]{martins2013building}
\bibinfo{author}{A.~C. Martins}, \bibinfo{author}{S.~Galam},
  \bibinfo{journal}{Physical Review E} \bibinfo{volume}{87}
  (\bibinfo{year}{2013}) \bibinfo{pages}{042807}.
\bibitem[{S{\^\i}rbu et~al.(2017)S{\^\i}rbu, Loreto, Servedio, and
  Tria}]{sirbu2017opinion}
\bibinfo{author}{A.~S{\^\i}rbu}, \bibinfo{author}{V.~Loreto},
  \bibinfo{author}{V.~D. Servedio}, \bibinfo{author}{F.~Tria}, in:
  \bibinfo{booktitle}{Participatory sensing, opinions and collective
  awareness}, \bibinfo{publisher}{Springer}, \bibinfo{year}{2017}, pp.
  \bibinfo{pages}{363--401}.
\bibitem[{Martins et~al.(2010)Martins, Pineda, and Toral}]{martins2010mass}
\bibinfo{author}{T.~V. Martins}, \bibinfo{author}{M.~Pineda},
  \bibinfo{author}{R.~Toral}, \bibinfo{journal}{EPL (Europhysics Letters)}
  \bibinfo{volume}{91} (\bibinfo{year}{2010}) \bibinfo{pages}{48003}.
\bibitem[{Carletti et~al.(2006)Carletti, Fanelli, Grolli, and
  Guarino}]{carletti2006make}
\bibinfo{author}{T.~Carletti}, \bibinfo{author}{D.~Fanelli},
  \bibinfo{author}{S.~Grolli}, \bibinfo{author}{A.~Guarino},
  \bibinfo{journal}{EPL (Europhysics Letters)} \bibinfo{volume}{74}
  (\bibinfo{year}{2006}) \bibinfo{pages}{222}.
\bibitem[{Quattrociocchi et~al.(2014)Quattrociocchi, Caldarelli, and
  Scala}]{quattrociocchi2014opinion}
\bibinfo{author}{W.~Quattrociocchi}, \bibinfo{author}{G.~Caldarelli},
  \bibinfo{author}{A.~Scala}, \bibinfo{journal}{Scientific reports}
  \bibinfo{volume}{4} (\bibinfo{year}{2014}) \bibinfo{pages}{4938}.
\bibitem[{Boschi et~al.(2020)Boschi, Cammarota, and Kühn}]{BOSCHI2020124909}
\bibinfo{author}{G.~Boschi}, \bibinfo{author}{C.~Cammarota},
  \bibinfo{author}{R.~Kühn}, \bibinfo{journal}{Physica A: Statistical
  Mechanics and its Applications} \bibinfo{volume}{558} (\bibinfo{year}{2020})
  \bibinfo{pages}{124909}.
\bibitem[{M{\"a}s et~al.(2010)M{\"a}s, Flache, and
  Helbing}]{mas2010individualization}
\bibinfo{author}{M.~M{\"a}s}, \bibinfo{author}{A.~Flache},
  \bibinfo{author}{D.~Helbing}, \bibinfo{journal}{PLoS Computational Biology}
  \bibinfo{volume}{6} (\bibinfo{year}{2010}) \bibinfo{pages}{e1000959}.
\bibitem[{M{\"a}s et~al.(2014)M{\"a}s, Flache, and Kitts}]{mas2014cultural}
\bibinfo{author}{M.~M{\"a}s}, \bibinfo{author}{A.~Flache},
  \bibinfo{author}{J.~A. Kitts}, in: \bibinfo{booktitle}{Perspectives on
  Culture and Agent-based Simulations}, \bibinfo{publisher}{Springer},
  \bibinfo{year}{2014}, pp. \bibinfo{pages}{71--90}.
\bibitem[{S{\^\i}rbu et~al.(2013)S{\^\i}rbu, Loreto, Servedio, and
  Tria}]{sirbu2013opinion}
\bibinfo{author}{A.~S{\^\i}rbu}, \bibinfo{author}{V.~Loreto},
  \bibinfo{author}{V.~D. Servedio}, \bibinfo{author}{F.~Tria},
  \bibinfo{journal}{Journal of Statistical Physics} \bibinfo{volume}{151}
  (\bibinfo{year}{2013}) \bibinfo{pages}{218--237}.
\bibitem[{Sznajd-Weron et~al.(2011)Sznajd-Weron, Tabiszewski, and
  Timpanaro}]{sznajd2011phase}
\bibinfo{author}{K.~Sznajd-Weron}, \bibinfo{author}{M.~Tabiszewski},
  \bibinfo{author}{A.~M. Timpanaro}, \bibinfo{journal}{EPL (Europhysics
  Letters)} \bibinfo{volume}{96} (\bibinfo{year}{2011}) \bibinfo{pages}{48002}.
\bibitem[{Radillo-D{\'\i}az et~al.(2009)Radillo-D{\'\i}az, P{\'e}rez, and del
  Castillo-Mussot}]{radillo2009axelrod}
\bibinfo{author}{A.~Radillo-D{\'\i}az}, \bibinfo{author}{L.~A. P{\'e}rez},
  \bibinfo{author}{M.~del Castillo-Mussot}, \bibinfo{journal}{Physical Review
  E} \bibinfo{volume}{80} (\bibinfo{year}{2009}) \bibinfo{pages}{066107}.
\bibitem[{Axelrod(1997)}]{axelrod1997dissemination}
\bibinfo{author}{R.~Axelrod}, \bibinfo{journal}{Journal of Conflict Resolution}
  \bibinfo{volume}{41} (\bibinfo{year}{1997}) \bibinfo{pages}{203--226}.
\bibitem[{Deffuant et~al.(2000)Deffuant, Neau, Amblard, and
  Weisbuch}]{deffuant2000mixing}
\bibinfo{author}{G.~Deffuant}, \bibinfo{author}{D.~Neau},
  \bibinfo{author}{F.~Amblard}, \bibinfo{author}{G.~Weisbuch},
  \bibinfo{journal}{Advances in Complex Systems} \bibinfo{volume}{3}
  (\bibinfo{year}{2000}) \bibinfo{pages}{87--98}.
\bibitem[{Macy et~al.(2003)Macy, Kitts, Flache, and
  Benard}]{macy2003polarization}
\bibinfo{author}{M.~W. Macy}, \bibinfo{author}{J.~A. Kitts},
  \bibinfo{author}{A.~Flache}, \bibinfo{author}{S.~Benard},
  \bibinfo{journal}{Dynamic Social Network Modeling and Analysis}
  (\bibinfo{year}{2003}) \bibinfo{pages}{162--173}.
\bibitem[{Flache and Macy(2011)}]{flache2011small}
\bibinfo{author}{A.~Flache}, \bibinfo{author}{M.~W. Macy},
  \bibinfo{journal}{The Journal of Mathematical Sociology} \bibinfo{volume}{35}
  (\bibinfo{year}{2011}) \bibinfo{pages}{146--176}.
\bibitem[{Mark(2003)}]{mark2003culture}
\bibinfo{author}{N.~P. Mark}, \bibinfo{journal}{American Sociological Review}
  (\bibinfo{year}{2003}) \bibinfo{pages}{319--345}.
\bibitem[{Huet and Deffuant(2010)}]{huet2010openness}
\bibinfo{author}{S.~Huet}, \bibinfo{author}{G.~Deffuant},
  \bibinfo{journal}{Advances in Complex Systems} \bibinfo{volume}{13}
  (\bibinfo{year}{2010}) \bibinfo{pages}{405--423}.
\bibitem[{Jager and Amblard(2005)}]{jager2005uniformity}
\bibinfo{author}{W.~Jager}, \bibinfo{author}{F.~Amblard},
  \bibinfo{journal}{Computational \& Mathematical Organization Theory}
  \bibinfo{volume}{10} (\bibinfo{year}{2005}) \bibinfo{pages}{295--303}.
\bibitem[{Hopfield(1982)}]{hopfield1982neural}
\bibinfo{author}{J.~J. Hopfield}, \bibinfo{journal}{Proceedings of the National
  Academy of Sciences} \bibinfo{volume}{79} (\bibinfo{year}{1982})
  \bibinfo{pages}{2554--2558}.
\bibitem[{Dall'Asta and Castellano(2007)}]{dall2007effective}
\bibinfo{author}{L.~Dall'Asta}, \bibinfo{author}{C.~Castellano},
  \bibinfo{journal}{EPL (Europhysics Letters)} \bibinfo{volume}{77}
  (\bibinfo{year}{2007}) \bibinfo{pages}{60005}.
\bibitem[{Jedrzejewski and Sznajd-Weron(2018)}]{jkedrzejewski2018impact}
\bibinfo{author}{A.~Jedrzejewski}, \bibinfo{author}{K.~Sznajd-Weron},
  \bibinfo{journal}{Physica A: Statistical Mechanics and its Applications}
  \bibinfo{volume}{505} (\bibinfo{year}{2018}) \bibinfo{pages}{306--315}.
\bibitem[{Stark et~al.(2008)Stark, Tessone, and
  Schweitzer}]{stark2008decelerating}
\bibinfo{author}{H.-U. Stark}, \bibinfo{author}{C.~J. Tessone},
  \bibinfo{author}{F.~Schweitzer}, \bibinfo{journal}{Physical review letters}
  \bibinfo{volume}{101} (\bibinfo{year}{2008}) \bibinfo{pages}{018701}.
\bibitem[{Garc{\'\i}a-Gavilanes et~al.(2017)Garc{\'\i}a-Gavilanes, Mollgaard,
  Tsvetkova, and Yasseri}]{garcia2017memory}
\bibinfo{author}{R.~Garc{\'\i}a-Gavilanes}, \bibinfo{author}{A.~Mollgaard},
  \bibinfo{author}{M.~Tsvetkova}, \bibinfo{author}{T.~Yasseri},
  \bibinfo{journal}{Science Advances} \bibinfo{volume}{3}
  (\bibinfo{year}{2017}) \bibinfo{pages}{e1602368}.
\bibitem[{Mariano et~al.(2020)Mariano, Mor{\u{a}}rescu, Postoyan, and
  Zaccarian}]{mariano2020hybrid}
\bibinfo{author}{S.~Mariano}, \bibinfo{author}{I.~Mor{\u{a}}rescu},
  \bibinfo{author}{R.~Postoyan}, \bibinfo{author}{L.~Zaccarian},
  \bibinfo{journal}{IEEE Control Systems Letters} \bibinfo{volume}{4}
  (\bibinfo{year}{2020}) \bibinfo{pages}{644--649}.
\bibitem[{Hebb(1949)}]{hebb1949organization}
\bibinfo{author}{D.~O. Hebb}, \bibinfo{title}{The organization of behavior: a
  neuropsychological theory}, \bibinfo{publisher}{J. Wiley; Chapman \& Hall},
  \bibinfo{year}{1949}.
\bibitem[{Amit et~al.(1987)Amit, Gutfreund, and
  Sompolinsky}]{amit1987statistical}
\bibinfo{author}{D.~J. Amit}, \bibinfo{author}{H.~Gutfreund},
  \bibinfo{author}{H.~Sompolinsky}, \bibinfo{journal}{Annals of Physics}
  \bibinfo{volume}{173} (\bibinfo{year}{1987}) \bibinfo{pages}{30--67}.
\bibitem[{Nadal et~al.(1986)Nadal, Toulouse, Changeux, and
  Dehaene}]{nadal1986networks}
\bibinfo{author}{J.~Nadal}, \bibinfo{author}{G.~Toulouse},
  \bibinfo{author}{J.~Changeux}, \bibinfo{author}{S.~Dehaene},
  \bibinfo{journal}{EPL (Europhysics Letters)} \bibinfo{volume}{1}
  (\bibinfo{year}{1986}) \bibinfo{pages}{535}.
\bibitem[{Parisi(1986)}]{parisi1986memory}
\bibinfo{author}{G.~Parisi}, \bibinfo{journal}{Journal of Physics A:
  Mathematical and General} \bibinfo{volume}{19} (\bibinfo{year}{1986})
  \bibinfo{pages}{L617}.
\bibitem[{van Hemmen et~al.(1988)van Hemmen, Keller, and
  K{\"u}hn}]{van1988forgetful}
\bibinfo{author}{J.~L. van Hemmen}, \bibinfo{author}{G.~Keller},
  \bibinfo{author}{R.~K{\"u}hn}, \bibinfo{journal}{EPL (Europhysics Letters)}
  \bibinfo{volume}{5} (\bibinfo{year}{1988}) \bibinfo{pages}{663}.
\bibitem[{Marinari(2019)}]{marinari2019forgetting}
\bibinfo{author}{E.~Marinari}, \bibinfo{journal}{Neural computation}
  \bibinfo{volume}{31} (\bibinfo{year}{2019}) \bibinfo{pages}{503--516}.
\bibitem[{Cohen and Grossberg(1983)}]{cohen1983absolute}
\bibinfo{author}{M.~A. Cohen}, \bibinfo{author}{S.~Grossberg},
  \bibinfo{journal}{IEEE transactions on systems, man, and cybernetics}
  (\bibinfo{year}{1983}) \bibinfo{pages}{815--826}.
\bibitem[{Hopfield(1984)}]{hopfield1984neurons}
\bibinfo{author}{J.~J. Hopfield}, \bibinfo{journal}{Proceedings of the National
  Academy of Sciences} \bibinfo{volume}{81} (\bibinfo{year}{1984})
  \bibinfo{pages}{3088--3092}.
\bibitem[{K{\"u}hn et~al.(1991)K{\"u}hn, B{\"o}s, and van
  Hemmen}]{kuhn1991statistical}
\bibinfo{author}{R.~K{\"u}hn}, \bibinfo{author}{S.~B{\"o}s},
  \bibinfo{author}{J.~L. van Hemmen}, \bibinfo{journal}{Physical Review A}
  \bibinfo{volume}{43} (\bibinfo{year}{1991}) \bibinfo{pages}{2084}.
\bibitem[{K\"uhn and B\"os(1993)}]{kuhn1993statistical}
\bibinfo{author}{R.~K\"uhn}, \bibinfo{author}{S.~B\"os},
  \bibinfo{journal}{Journal of Physics A: Mathematical and General}
  \bibinfo{volume}{26} (\bibinfo{year}{1993}) \bibinfo{pages}{831}.

\end{thebibliography}

\end{document}